\documentclass{iopart}             
\usepackage{iopams}
\usepackage{graphicx}
\usepackage{url,graphics,epsfig}
\usepackage{amssymb}
\usepackage[breaklinks=true,colorlinks=true,linkcolor=blue,urlcolor=blue,citecolor=blue]{hyperref}
\usepackage{cite}
 \usepackage{longtable}

\begin{document}

\jl{2}
%
%
%
\def\etal{{\it et al~}}
%
%
%
%
%
%
\setlength{\arraycolsep}{2.5pt}             

\title[Photoionisation of Ca$^+$ ions in the valence region]{Photoionisation of Ca$^{+}$ ions
       in the valence-energy region 20~eV~--~56~eV : experiment and theory}

\author{A~M\"{u}ller$^{1}\footnote[1]{Corresponding author, E-mail: Alfred.Mueller@iamp.physik.uni-giessen.de}$,
	  S~Schippers$^{1,2}$, R~A~Phaneuf$^3$,  A~M~Covington$^3$,   A~Aguilar $^{3}\footnote[3]{Present address: ReVera Incorporated,
                                                          3090 Oakmead Village Drive, Santa Clara, CA 95051, USA}$,
                                                          G~Hinojosa$^{3,4}$,
            J~Bozek$^{5}\footnote[2]{Present address: Synchrotron SOLEIL - L'Orme des Merisiers,
                                        Saint-Aubin - BP 48 91192 Gif-sur-Yvette cedex, France}$,
	 M~M~Sant'Anna$^{5,6}$, A~S~Schlachter$^{5}$, C~Cisneros$^{4}$
              and B M McLaughlin$^{7,8}\footnote[4]{Corresponding author, E-mail: bmclaughlin899@btinternet.com}$}

\address{$^{1}$Institut f\"{u}r Atom- ~und Molek\"{u}lphysik,
                         Justus-Liebig-Universit\"{a}t Gie{\ss}en, 35392 Giessen, Germany}

\address{$^{2}$I. Physikalisches Institut,
                           Justus-Liebig-Universit\"{a}t Gie{\ss}en, 35392 Giessen, Germany}

\address{$^{3}$Department of Physics, University of Nevada,
                          Reno, NV 89557, USA}

\address{$^{4}$Instituto de Ciencias F\'isicas, Universidad Nacional Aut\'onoma de M\'exico,
                           Cuernavaca 62251, Mexico}

\address{$^{5}$Advanced Light Source, Lawrence Berkeley National Laboratory,
                          Berkeley, California 94720, USA }

 \address{$^{6}$Instituto de F\'\i sica, Universidade Federal
                              do Rio de Janeiro, 
                              21941-972 Rio de Janeiro RJ, Brazil}

\address{$^{7}$Department of Physics and Astronomy, Center for Simulational Physics,
                          University of Georgia,
                          Athens, GA 30602-2451, USA}

\address{$^{8}$Centre for Theoretical Atomic, Molecular and Optical Physics (CTAMOP),
                          School of Mathematics and Physics,
                          Queen's University Belfast, Belfast BT7 1NN, UK}

%
%
\begin{abstract}
Relative cross sections for the valence shell photoionisation (PI)
of $\rm ^2S$ ground level and $\rm ^2D$ metastable Ca$^{+}$ ions
were measured with high energy resolution by using the ion-photon merged-beams technique at the
Advanced Light Source. Overview measurements were
performed with a full width at half maximum bandpass of $\Delta E =17$~meV,
covering the energy range 20~eV -- 56~eV. Details of the PI spectrum were
investigated at energy resolutions reaching the level of $\Delta E=3.3$~meV. The photon energy scale
was calibrated with an uncertainty of $\pm5$~meV.
By comparison with previous absolute measurements
the present experimental high-resolution data were
normalised to an absolute cross-section scale and the fraction of metastable Ca$^{+}$ ions
that were present in the parent ion beam was determined to be 18$\pm$4\%.
Large-scale {\textit R}-matrix calculations using the Dirac Coulomb approximation
and employing 594 levels in the close-coupling expansion were performed for the
Ca$^{+}(3s^23p^64s~^2\textrm{S}_{1/2})$ and Ca$^{+}(3s^2 3p^6 3d~^2\textrm{D}_{3/2,5/2})$ levels.
The experimental data are compared with the results of these calculations and previous
theoretical and experimental studies.
\end{abstract}


\vspace{1.0cm}
\begin{flushleft}
Short title: Photoionisation of Ca$^{+}$ ions  in the valence region\\
\vspace{0.25cm}
J. Phys. B: At. Mol. Opt. Phys. \today
\end{flushleft}


\maketitle

%
%
%

\section{Introduction}

Multiply excited states play an important role in electron-ion and photon-ion interactions; cross sections are often dominated by indirect processes, forming intermediate autoionizing states which then decay into the different final channels that are observed in various types of experiments. In particular, autoionizing states are most relevant in photoionisation (PI) and its time-reversed analog, photorecombination (PR) of ions, where they produce resonances superimposed on smooth cross sections for direct reaction channels. The pathways of a reaction via direct and resonance channels may not always be distinguishable. If the initial and final states are identical, the amplitudes of the pathways interfere and, as a result, the associated cross sections show  resonance contributions with distorted line shapes, the so called Fano profiles ($\!$\cite{Mueller2009b} and references therein). Studying such interference in detail can provide deep insights into the  quantum dynamics of atomic-scale systems~\cite{Kaldun2017}.

While Fano line shapes are commonly observed in PI, electron-ion recombination experiments rarely show direct evidence for interference phenomena. Because of the importance of such effects, theoretical attempts have been made to predict interference of resonant (indirect) and direct PR channels as well as interference of interacting resonances~\cite{Connerade1988,Labzowsky1994}. A particular example is the prediction of an asymmetric resonance profile in PR of  Sc$^{3+}$ ions by Gorczyca \etal~\cite{Gorczyca1997} involving the following transitions
 \begin{eqnarray}
 \label{Eq:PR}
 \textrm{e}^- + (3s^2 3p^6~^1\textrm{S}) &~ \to ~ &
 (3s^2 3p^5 3d^2~^2\textrm{F})\nonumber \\
  ~ & \searrow   & \downarrow\\
  ~ &~ & (3s^2 3p^6 3d~^2\textrm{D}) + \gamma\nonumber.
\end{eqnarray}
The two indistinguishable pathways to the final $^2\textrm{D}$ term with a photon $\gamma$ emitted produce an asymmetric resonance line shape, and the fact that the intermediate $^2\textrm{F}$ term can decay by a very fast Super-Koster-Cronig transition makes this resonance extremely broad, facilitating its detailed mapping in a recombination experiment.

 This carefully chosen example set the starting point for a very fruitful research initiative on PR of argon-like and PI of potassium-like ions both in theory and experiment aiming at the detailed spectroscopy of multiply excited levels of potassium-like ions in low charge states. First experiments on the recombination of Ar-like Sc$^{3+}$  ions~\cite{Schippers1998b,Schippers1999a} did not show the predicted asymmetric resonance in the expected energy range, indicating the difficulties to correctly predict level energies in an atomic system where the $3d$ and $4s$ subshells compete for providing the lowest binding energy for an electron and where an additional $3p$ vacancy is present. Therefore, experiments were started on PR of Ti$^{4+}$~\cite{Schippers1998a} which is isoelectronic with Sc$^{3+}$. It became clear that the prediction of resonance energies, i.e., energies of autoionizing levels in K-like ions, and of their decay properties  is much more challenging than previously anticipated. New measurements on PR of Sc$^{3+}$ ions~\cite{Schippers2002a}, conducted under greatly improved experimental conditions, clearly revealed the existing discrepancies between calculated and observed resonance energies and cross sections. Motivated by this unsatisfying situation the experimental efforts were extended to PI of potassium-like Sc$^{2+}$~\cite{Schippers2002b,Schippers2003a} and Ti$^{3+}$ ions ~\cite{Schippers2004b} and by exploiting the principle of detailed balance based on time-reversal symmetry, the measurements on PR and PI provided a very consistent picture confirming the initial prediction of asymmetric resonances as a consequence of interference of resonant and direct interaction channels. Within this same spectroscopy-oriented initiative, experiments were conducted on PI of potassium-like Ca$^+$, for which preliminary results were reported~\cite{Mueller2001a},  and on PI of Fe$^{7+}$ ions~\cite{Gharaibeh2011a}.

Substantial progress was achieved on the theory side in the treatment of PR of argon-like ions and PI of potassium-like ions~\cite{Sossah2008,Nikolic2009,Nikolic2010a,Sossah2010a,Sossah2012,Wang2015}. The PI and PR spectra of scandium and titanium ions of relevance in the present context appear to now be well understood. However, the existing calculations for PI of Fe$^{7+}$ ions~\cite{Sossah2010a,Tayal2015} do not reproduce the experimental findings~\cite{Gharaibeh2011a} in a satisfying manner, and the reason for this is not yet clear. The situation for Ca$^+$, the lightest ion in the potassium isoelectronic sequence,  is the topic of the present investigation. Complications in the theoretical treatment are to be expected because of the low parent-ion charge state and the corresponding strong influence of electronic correlations and inter-channel coupling which give rise to the subtle interplay of the $3d$ and $4s$ subshells.

The detailed investigation and understanding of electronic structure observable in PI of Ca$^+$, and consequently also in PR of Ca$^{2+}$ ions, is not the only motivation for research on calcium ions in low charge states.
Properties of the Ca$^{+}$ ion are of great interest
and  importance in astrophysics.
For example, the absorption spectrum of the Ca$^{+}$ ion in its $4s~^2\textrm{S}$ ground level or its long-lived excited $3d~^2\textrm{D}$ levels is used to
explore the structure and properties of interstellar
dust clouds~\cite{Hobbs1988,Welty1996}. The  interstellar lines
of Ca II (IS CaII K and IS Ca II H)  have  been
observed in the spectrum of the bright Seyfert galaxy NGC 3783~\cite{West1985}.
Emission in the Ca II H  and K line cores has long been
known to be a good proxy for magnetic activity in the Sun~\cite{Hall2008}.
PI of Ca$^{+}$(Ca II) ions
out of the metastable excited $3d~^2\textrm{D}$ term is important for understanding the Sun
and sunspot atmospheres~\cite{Linsky1970,Rowe1992}.
Ca II H spectra~\cite{Beck2005, Beck2012}
have been used to examine whether spicules can be detected
in the broad-band Ca~II H filter imaging data on the solar disc~\cite{Beck2013b}.
Spectral lines of Ca~II  have also been observed  in the photospheric spectra of  luminous supernovae~\cite{GalYam2012}.
Fundamental atomic data on Ca II  are required to understand all these observed spectra.

Few measurements have been made on the Ca$^{+}$ ion.
Spectroscopic measurements on photoabsorption by Ca, Ca$^*$, and Ca$^+$ in the range of excitations of a $3p$ electron were first reported by Sonntag~\etal~\cite{Sonntag1986} who produced a plasma by laser-irradiation of Ca vapour and observed the absorption of quasi-continuum VUV backlight emitted by a second laser-generated plasma.
Absolute photoionisation cross sections
were first measured by Lyon \etal~\cite{Lyon1987a}, who merged beams of tunable monochromatic VUV synchrotron radiation
(282 \AA~ $\le \lambda \le$~ 498 \AA) and ground-level Ca$^{+}$ ions.
Electron spectroscopy experiments were performed on Ca$^{+}$  by Bizau~\etal~\cite{Bizau1991} and a similar attempt was undertaken on metastable
Ca$^{+}$ ions by Gottwald~\etal~\cite{Gottwald1997}.
Valence-shell PI studies on the
ground and metastable levels of Ca$^{+}$ were
carried out by Kjeldsen \etal~\cite{Kjeldsen2002d} in the
photon energy range 28.0 -- 30.5~eV.  PI along the sequence of ions from Ca$^+$ to Ni$^+$
in the range of $ 3p \rightarrow 3d$ transitions was studied experimentally by  Hansen \etal~\cite{Hansen2007a}.

Diverse theoretical
methods have been used  to study PI of  neutral and singly ionized calcium.
Altun and Kelly~\cite{Altun1985} used many-body perturbation
theory (MBPT) to study PI of the neutral atom
with excitation to the residual Ca$^{+}$ $ 3d$ and $ 4p$ levels.
For the Ca$^{+}$ ion the first {\it ab initio} calculations using the
{\textit R}-matrix method in an $LS$-coupling scheme were carried out
 by Miecznik \etal~\cite{Miecznik1990} for the energy range from above the first ionisation
 threshold to a photon energy of 43.54~eV.
Configuration-interaction wavefunctions generated from the
$ 3s^2 3p^5n\ell$ ($n\ell$ = $ 4s$, $ 4p$, $ 3d$ and
$\overline{4d}$), where $\overline{ 4d}$ is a pseudo-orbital,
were included in the {\textit R}-matrix close-coupling expansion.
The corresponding Rydberg series of resonances
were analyzed and compared with the early experimental
data of Lyon \etal \cite{Lyon1987a}.
The $LS$-coupling scheme used by Miecznik \etal in their
pioneering work on Ca$^+$ neglected relativistic effects.

Following the work of Miecznik~\etal~\cite{Miecznik1990},
calculations by Ivanov and West~\cite{Ivanov1993}, using
the non-relativistic, spin polarized version of the
random-phase approximation with exchange (RPAE), determined positions and intensities of the resonances seen
experimentally in the PI spectrum of Ca$^{+}$
in the range 27 -- 43~eV~\cite{Lyon1987a}. Values for the oscillator strengths
and energy positions for the main one-electron transitions were
obtained, and this information was used to tentatively identify
 the most prominent features in the experimental spectrum.
 Hansen and Quinet carried out calculations on $3p$ subshell excitation from metastable Ca$^+$ ions using the Cowan code~\cite {Cowan1981} with 26 interacting configurations. They provided resonance energies for the 26 strongest  $3p^6 3d \to 3p^5 3d^2$ transitions as well as the associated oscillator strengths.
Large-scale configuration interaction calculations for ground-level Ca$^+$ in intermediate coupling
 by Hibbert and Hansen \cite{Hibbert1999} using a total of 206 configurations provided recommended level assignments for the spectrum obtained by Lyon~\etal together with resonance energies and oscillator strengths.
Many-body perturbation theory was used by Jiang~\etal~\cite{Jiang2001} to calculate the resonance structure of $3p \to 3d$ transitions in ground-level Ca$^+$ ions.

More recently, PI cross-section calculations for Ca$^+$ were  extended
  by Sossah \etal  \cite{Sossah2012}
using the {\textit R}-matrix method in both
$LS$ and intermediate coupling ($lsj$) approximations.
Configuration interaction wavefunctions, generated from
 $ 3s^2 3p^5n\ell$, ($n\ell$ = $ 4s$, $ 4p$,
 $ 3d$,  $\overline{ 4d}$,  $\overline{ 5s}$, $\overline{ 5p}$
 and  $\overline{ 5d}$),
 where $\overline{ 4d}, \dots \overline{ 5d}$
 are pseudo-orbitals, were included in the {\textit R}-matrix expansion.
The work of Sossah \etal \cite{Sossah2012} was restricted to including 17 fine-structure
 levels in  the close-coupling calculations, where the residual
 states were represented by highly sophisticated correlated configuration wavefunctions
generated from selective one and two-electron promotions.
 The results of Sossah \etal \cite{Sossah2012}  were in very respectable
 agreement with the previous experimental work performed by Kjeldsen~\etal~\cite{Kjeldsen2002d}.

The availability of large-scale computational facilities now makes it possible to extend previous theoretical  work by using close-coupling
calculations within the Dirac Coulumb ($jj$) {\textit R}-matrix approximation with very large basis sets.
Corresponding results are compared with previous experimental studies and with the present measurements, which were performed over a wider energy range and  at higher energy resolution as compared to previous experiments using undulator radiation from a third-generation synchrotron source.

The layout of this paper is as follows.  Section 2 presents  the experimental procedures  and
section 3 outlines the corresponding theoretical approach.
In section 4 the results from the present findings are discussed.  Finally in section 5  the
results of this work are summarised.

\section{Experiment}\label{sec:experiment}

The experiments on PI of Ca$^{+}$  ions made use of the
Ion-Photon Beam (IPB) endstation of undulator beamline 10.0.1.2 at the
Advanced Light Source (ALS) in Berkeley, California, USA. The merged-beams technique~\cite{Phaneuf1999,Schippers2016} was employed. Experimental methods
for studying photoabsorption by ions and typical results of such experiments have
been reviewed by M\"{u}ller \etal~\cite{Mueller2015b,Mueller2015c}. The most recent publication of experimental results from the IPB addressed PI of W$^{4+}$  ions~\cite{Mueller2017b}.
The general layout of the  IPB setup and the associated experimental procedures were described in detail by Covington \etal~\cite{Covington2002a}. Here, only a brief overview of the experiment is presented with details specific to the present measurements.

Beams of Ca$^+$ ions of 6~keV energy  were produced using a commercial hot-filament, low-pressure-discharge ion source on a positive potential of 6~kV. Metallic calcium was deposited inside the source body and evaporated into the discharge plasma. The source had been developed on the basis of the electron-bombardement plasma source described by Menzinger and W{\aa}hlin~\cite{Menzinger1969}. This type of ion source has often been employed in experiments where pure beams of ions in their ground level were desired.  For producing such beams the anode voltage in the source is kept sufficiently low that electrons do not gain enough energy to reach excited ionic levels from the ground state of the neutral atom. In contrast,  the anode voltage was purposely set at a relatively high voltage (80~V) in the present experiment. The associated electron energy of 80~eV is far above the threshold for the production of the Ca$^+(3p^6 3d~^2{\textrm D})$ excited term.  This was desired  for the determination of  the energy and the width of the doubly excited $3p^5 3d^2~^2{\textrm F}$ levels (see equation~\ref{Eq:PR}) that can be populated by photoexcitation of Ca$^+(3p^6 3d~^2{\textrm D})$ ions. The lifetimes of Ca$^+(3p^6 3d~^2{\textrm D}_{3/2})$ and Ca$^+(3p^6 3d~^2{\textrm D}_{5/2})$ have been determined to be 1195(8)~ms~\cite{Shao2016} and 1168(7)~ms~\cite{Barton2000}, respectively. These lifetimes are more than four orders of magnitude longer than the flight time of the Ca$^+$ ions from the source to the photon-ion interaction region. Collisional quenching of the metastable levels by interactions with the residual gas can be neglected in the high to ultra-high vacuum of the apparatus. Hence, an initial population of the long-lived excited levels does not change while the ions travel to the interaction region. Analysis of the data taken in the experiment showed that 18\% of the ions in the primary beam were in the $3p^6 3d~^2{\textrm D}$ metastable term (for details see below).

The ion beam extracted from the source was analysed with respect to mass per charge by a 60$^\circ$ dipole bending magnet. The selected $^{40}$Ca$^+$ beam component was collimated and transported by suitable electrostatic lens and steering elements to an electrostatic 90$^\circ$ spherical deflector, the \textit{merger},  which directed the parent ion beam  onto the axis of a monoenergetic beam of photons from the beamline monochromator. In the merging section between the 90$^\circ$ spherical deflector and a subsequent 45$^\circ$ bending magnet, the \textit{demerger}, parent $^{40}$Ca$^+$ ions were excited and ionized by photons. After  ion charge-state separation by the demerger magnet the parent beam was collected in a Faraday cup and the Ca$^{2+}$ product ions were detected by a microsphere-plate detector. The photon flux was measured by a calibrated Si X-ray diode. The photon energy was scanned in steps of 5~meV for the overview spectrum covering the range from 20 to 56~eV at 17~meV resolution. The step size was reduced in narrower energy ranges when smaller bandwidths of photon energies were employed (e.g. 0.5~meV steps for 4.6~meV resolution).  At each photon energy, data were alternately taken for 1~s with both beams ''on''  and then again for 1~s  with the photon beam ''off''. The integrated ion current, the photon flux, and the number of  product Ca$^{2+}$ ions collected in each of the ''on'' and ''off'' phases were recorded, permitting subtraction of the detector background which arises mainly from collisions of the parent ions with the residual gas and from stray particles and photons produced when the ions hit surfaces.

In one energy scan a range of typically 1~eV was covered assuming that the form factor\cite{Phaneuf1999}, which quantifies the overlap of the photon beam and the ion beam,  does not significantly change over such a short energy interval. For checking the reproducibility and consistency of the measurements, additional scans were recorded in energy ranges spanning up to 5.2 eV. Numerous overlapping scans were measured to cover the total energy range of interest.  The resulting  overview  spectrum at 17$\pm$2~meV resolution comprises more than 7200 data points. In addition, shorter energy ranges were scanned at higher resolution, including bandpasses of 8.8~meV, 4.6~meV, and 3.3~meV. Because of the availability of absolute cross sections for PI of Ca$^+$ in smaller energy regions and at lower resolving powers~\cite{Lyon1987a,Kjeldsen2002d} no attempt was made to carry out absolute measurements in the present study. The focus was rather on  well calibrated photon energies and high-resolution spectroscopy of autoionizing levels in the Ca$^+$ ion.

For the determination of the ion-rest-frame photon energy the Doppler effect resulting from the counter-propagation of  photons and the 1.7$\times 10^7$~cm/s Ca$^+$ ions was taken into account. These corrections were between 10 and 30~meV in the investigated energy range. The photon energy was calibrated by an ionisation-threshold measurement with 6~keV K$^+$ ions (31.62500(19)~eV~\cite{NIST2016}) and the observation of $3p^6 4s \to 3p^5 4s^2~^2{\textrm P}$ resonances in the Ca$^+$ ion ($3p^5 4s^2~^2{\textrm P}_{3/2}$ at 28.1995~eV and $3p^5 4s^2~^2{\textrm P}_{1/2}$ at 28.5441~eV~\cite{NIST2016}). The total absolute uncertainty of the energy scale is estimated to be $\pm $5~meV. In the determination of specific resonance energies from the scan measurements the statistical uncertainties of the measured cross sections can introduce additional uncertainties.

The raw spectrum obtained in the experiment was corrected for effects of higher-order radiation known to be present at the ALS photon beamline employed in the present measurements. These effects depend on the photon energy and are significant below approximately 25~eV. The necessary corrections have been discussed previously by M\"{u}ller~\etal~\cite{Mueller2015h}. With these corrections the ion yield was normalised to ion current and photon flux, i. e., the relative cross section function for PI of Ca$^+$ ions was obtained.

The present relative PI cross sections were normalised to the previous absolute measurements by Lyon \etal and Kjeldsen \etal$\!$. In a first step, the overview scan spectrum was normalized to the data of Lyon \etal which cover a relatively wide range of photon energies from 28 to 42.5~eV. Lyon \etal used a surface ionisation source that produced only ground-level Ca$^+(3p^6 4s~^2{\textrm S}_{1/2})$ ions. Hence, the normalized scan spectrum which was taken with a mixture of ground-state and metastable Ca$^+$ ions was still to be corrected for the metastable-ion fraction in the parent ion beam.  Kjeldsen \etal were able to produce mixed beams of Ca$^+$ ions in the $^2\textrm S$ ground-level and $^2\textrm D$ metastable ions with different fractional compositions. From the measured absolute cross sections for different mixtures they were able to infer the absolute cross sections $\sigma_g$ for ground-level Ca$^+(^2{\textrm S})$ and $\sigma_m$ for metastable Ca$^+(^2{\textrm D})$. The complete resonance strength contained in the measured spectrum $\sigma_g$ in the energy range 28.1 --30.1~eV is 20.6~Mb\,eV in the Kjeldsen \etal data and 23.1~Mb\,eV in the Lyon \etal measurement. The two numbers differ by about 11\%, well within the combined error bars of the two data sets.  In the present experiment a mixed beam was used with the fraction $m$ of metastable ions not known {\textit a priori}. Thus, the measured relative cross section $\sigma_{rel}$ can be described as
\begin{equation}
\label{Eq:apparentxsection}
\sigma_{rel} = \alpha [m\sigma_m + (1-m)\sigma_g]
\end{equation}
where $\alpha$ is an overall calibration factor.
Fortunately, most of the resonances observed in the spectrum $\sigma_{rel}(E_{ph})$ in the energy range 28 -- 30.5~eV could be unambiguously assigned to either the ground-level component or the metastable fraction of the ion beam. Only few resonances are not sufficiently well resolved to make such an assignment. The cross sections $\sigma_{rel}$, $\sigma_g$, and $\sigma_m$ were integrated over distinct energy regions $\epsilon_1$ and $\epsilon_2$ containing only ground-level resonances and only metastable-level resonances, respectively. The resulting resonance strengths are denoted $S_{rel,g}$, $S_{rel,m}$, $S_{g}$, and $S_{m}$. According to equation~\ref{Eq:apparentxsection}, the following relations hold
\begin{eqnarray}
\label{Eq:normalizationg}
S_{rel,g} = \alpha (1-m) S_{g}\\
\label{Eq:normalizationm}
S_{rel,m} = \alpha    m  S_{m}.
\end{eqnarray}
The ratio of the two equations yields the fraction $m$ of metastable ions in the parent ion beam
\begin{equation}
\label{Eq:fraction}
m = S_{rel,m} S_{g}  (S_{rel,m} S_{g} + S_{rel,g} S_{m})^{-1}.
\end{equation}
Once $m$ is known, $\alpha$ immediately follows from either one of the equations~\ref{Eq:normalizationg} or~\ref{Eq:normalizationm} and hence the normalised cross section for the mixed beam $\sigma_{mix} = \sigma_{rel} / \alpha$. If one considers that the cross sections measured by Lyon~\etal are 11\% higher than those of Kjeldsen~\etal it seems justified to introduce an additional average correction factor of 1.055. In the following the cross section measured for a mixed beam of metastable and ground-level ions in the present study is called $\sigma$. It follows from $\sigma_{mix}$ by multiplication of the correction factor 1.055 which essentially normalises the present measurement to the average of the results obtained by Lyon \etal and Kjeldsen \etal$\!$. The analysis of the measured PI spectrum and the data provided by Kjeldsen \etal yielded a fraction $m=0.18 \pm 0.04$ of metastable ions in the parent ion beam used in the present experiment.

The uncertainties of $\sigma_g$ and $\sigma_m$ quoted by Kjeldsen \etal are 10-15\% and 25\%, respectively. This results in an uncertainty of 0.04 for $m$. The systematic uncertainty of the present cross section $\sigma$ is estimated at about 30\%. This number results from the combined uncertainties of the Aarhus measurements and the normalization procedure itself. Comparison of peak areas of $3p^5 4s nd~^2{\textrm P})$ resonances associated exclusively with ground-level parent Ca$^+$ ions and resolved in both the present and the  Lyon-\etal spectra shows agreement within a factor $0.82 \pm 0.15$. This factor has to be expected for a 82\% fraction  of ground-level ions in the present experiment indicating that 30\% is a conservative estimate of the relative systematic uncertainty.

\section{Theory}\label{sec:theory}

 High-resolution PI cross-section measurements require state-of-the-art theoretical methods
with relativistic effects \cite{Grant2007}, in order  to obtain suitable
agreement with experiment. This has been demonstrated for the Ca$^+$ ion by Sossah \etal~\cite{Sossah2012} who compared nonrelativistic and relativistic (Breit-Pauli) {\textit R}-matrix calculations with the experimental results available at the time.
 The present work employs an efficient parallel version~\cite{Ballance2006,Fivet2012}  of the
 Dirac-Atomic {\textit R}-matrix-Codes (DARC) \cite{Norrington1987,Wijesundera1991,darc,McLaughlin2012a,McLaughlin2012b} developed for treating
 electron and photon interactions with atomic systems.  This suite continues to evolve
\cite{McLaughlin2015a,McLaughlin2015b,McLaughlin2016c,McLaughlin2017b}
in order to provide for ever increasing larger expansions of target
and collision models for electron and photon impact with heavy atomic systems.

The DARC suite of codes
allows  high-quality PI cross section  calculations to
be performed on heavy complex systems.
PI cross-section calculations of Se$^+$~\cite{McLaughlin2012b}, Se$^{2+}$~\cite{Macaluso2015},
 Xe$^+$~\cite{McLaughlin2012a}, Kr$^+$~\cite{McLaughlin2012a,Hinojosa2012},
 Xe$^{7+}$~\cite{Mueller2014b},  $2p^{-1}$ inner-shell
 studies on Si$^+$ ions~\cite{Kennedy2014},
Ar$^{+}$~\cite{Tyndall2016a}, and Co$^+$~\cite{Tyndall2016b},
valence-shell studies on
neutral sulfur~\cite{Barthel2015}, sulfur-like chlorine, Cl$^{+}$~\cite{McLaughlin2017c}, copper-like zinc, Zn$^+$~\cite{Hinojosa2017},
tungsten and its ions, W~\cite{Ballance2015a}, W$^{+}$  \cite{Mueller2015h},
W$^{2+}$, W$^{3+}$  \cite{McLaughlin2016a}, W$^{4+}$ \cite{Mueller2017b} and W$^{63+}$ \cite{Turkington2016}
 have been made using these DARC codes.  Suitable agreement of the DARC
photoionisation cross-sections with high resolution measurements
performed at leading synchrotron light sources was obtained.

%
%
\begin{table}
\begin{center}
\caption{\label{tab1}Comparison of  theoretical energies of the Ca$^{2+}$
	    ion with the NIST~\cite{NIST2016} tabulated data.
        A sample of the lowest 16  NIST levels of the Ca$^{2+}$
	    ion from the present large-scale GRASP calculations~\cite{Dyall1989,Parpia2006,Grant2007}
	    is compared with previous theoretical work
	    (AUTOS and MCHF~\cite{Froese-Fischer2006a,Nikolic2010a,Sossah2012}).
	    Energies are given in Rydbergs (Ry) relative to the ground state
	    of the Ca$^{2+}$ ion.}
\scriptsize
 \lineup
\item[]\begin{tabular}{ccccccccccc}
\br
Level       	&  State 			&  Term	 	& NIST$^a$ 	&GRASP$^b$   & GRASP$^{c}$	&GRASP$^{d}$  &AUTOS$^e$ &AUTOS$^f$ &MCHF$^g$ \\
		&				&	            		&(Ry)   	&  (Ry)      	& (Ry)		& (Ry)		& (Ry)	 	& (Ry)    	& (Ry)	 \\	
\mr
 1  		& $3p^6$ 	&  $\rm ^1S_{0}$         		&0.00000          &0.00000    &0.00000     &0.00000       &0.00000       &0.00000       &0.00000\\
 2  		& $3p^53d$	&  $\rm ^3P^o_{0}$   	 	     &1.85327         &2.00695    &1.82269     &1.85481       &1.84254       &1.87953       &1.76015\\
 3  		& $3p^53d$	&  $\rm ^3P^o_{1}$    	          &1.85764        &2.01158    &1.82740     &1.85929       &1.84754       &1.88428       &1.76449\\
 4  		& $3p^53d$	&  $\rm ^3P^o_{2}$    	          &1.86666        &2.02106    &1.83702     &1.86885       &1.85760       &1.89391       &1.77311\\
 5  		& $3p^53d$	&  $\rm ^3F^o_{4}$    	          &1.93471        &2.09287    &1.90927      &1.93933       &1.94250       &1.95288       &1.84398\\
 6  		& $3p^53d$	&  $\rm ^3F^o_{3}$    	          &1.94446        &2.10324    &1.91964      &1.94970       &1.95234       &1.96321       &1.85310\\
 7  		& $3p^53d$	&  $\rm ^3F^o_{2}$    	          &1.95316        &2.11228    &1.92875      &1.95874       &1.96138       &1.97260       &1.86119\\
 8  		& $3p^53d$	&  $\rm ^1D^o_{2}$    	          &2.05788        &2.24472    &2.06612      &2.09184       &2.07332       &2.08651       &1.97206\\
 9  		& $3p^53d$	&  $\rm ^3D^o_{3}$   	          &2.06250        &2.24840    &2.06274      &2.08817       &2.07134       &2.08060       &1.97626\\
 10  		& $3p^53d$	&  $\rm ^3D^o_{1}$    	          &2.07251        &2.25538    &2.07303      &2.09861       &2.08111       &2.09323       &1.98530\\
 11  		& $3p^53d$	&  $\rm ^3D^o_{2}$   	          &2.07211        &2.25678    &2.07472      &2.10030       &2.08280       &2.09523       &1.98484\\
 12 		& $3p^53d$	&  $\rm ^1F^o_{3}$    	          &2.08146        &2.26471    &2.08347      &2.10831       &2.09192       &2.08146       &1.99576\\
 13 		& $3p^54s$	&  $\rm ^3P^o_{2}$    	          &2.21025        &2.36114    &2.20408      &2.19121       &2.19563       &2.31306       &2.11402\\
 14 		& $3p^54s$	&  $\rm ^3P^o_{1}$    	          &2.22286        &2.37496    &2.21841      &2.20474       &2.21091       &2.32712       &2.12631\\
 15 		& $3p^54s$	&  $\rm ^3P^o_{0}$   	          &2.23818        &2.39076    &2.23414      &2.22083       &2.22495       &2.34139       &2.14026\\
 16 		& $3p^54s$	&  $\rm ^1P^o_{1}$   	          &2.25718        &2.41325    &2.25854      &2.24141       &2.25920       &2.36902       &2.16055\\
\mr
\end{tabular}
\\
\begin{flushleft}
$^{a}$NIST tabulations \cite{NIST2016}\\
$^{b}$GRASP, present 93-levels treatment\\
$^{c}$GRASP, present 594-levels treatment \\
$^{d}$GRASP,  present 1153-levels treatment\\
$^{e}$AUTOS, Sossah et al  \cite{Sossah2012}\\
$^{f}$AUTOS, Nikoli\'c et al \cite{Nikolic2010a}\\
$^{g}$MCHF, Froese Fischer et al \cite{Froese-Fischer2006a}.
\end{flushleft}
\end{center}
\end{table}

For PI cross-section calculations on the Ca$^{+}$ ion,
close-coupling DARC computations were initially performed using 93 levels of the residual
Ca$^{2+}$ ion arising from single electron promotion to the
$n$=3, 4 and 5 levels from the ground-state configuration,   $3s^23p^6$, resulting in a total of eleven configurations.
These configurations  were
$3s^23p^6$,  $3s^23p^54s$,  $3s^23p^53d$,
$3s^23p^54p$, $3s^23p^54d$, $3s^23p^54f$,
$3s^23p^55s$, $3s^23p^55p$, $3s^23p^55d$,
$3s^23p^55f$.  In addition,  effects of opening the
3s shell were included by adding the single configuration $3s3p^63d$ to the basis set.

Next, a model was used that included one-electron promotions $3p \rightarrow (3d, \dots, 5\ell)$,  where $\ell=s, p, d, f$,
and  two-electron promotions $3s^23p^6 \rightarrow 3s^23p^4(n\ell)^2$
to the $n$ = 3 and 4 levels together with $3s^23p^6 \rightarrow 3s^23p^43d4s$.
In order to have compact target and scattering representations,  it was necessary to restrict
the two-electron configurations to
$3s^23p^43d^2$,  $3s^23p^44s^2$,
$3s^23p^44p^2$, $3s^23p^44d^2$,
$3s^23p^44f^2$, and $3s^23p^43d4s$,
giving a total of 594 levels.

A final target model incorporating
one-electron promotions $3p \rightarrow (3d, \dots, 5\ell)$ where $\ell=s, p, d, f$ and
 the two-electron promotions
$3p^2 \rightarrow (3d4\ell, \dots 3d5\ell)$
and $3p^2 \rightarrow (3d^2, 4 \ell^2, \dots, 5\ell^2)$, giving 1153 levels
was also investigated.

Table \ref{tab1} shows a sample of results from selective models for the energy levels
using the GRASP code for the lowest 16 levels of the  residual Ca$^{2+}$  ion.
The present results are compared with previous theoretical studies from
the AUTOSTRUCTURE (AUTOS) \cite{Nikolic2010a,Sossah2012}  and MCHF codes \cite{Froese-Fischer2006a}
and with the NIST tabulated values \cite{NIST2016}. From the results presented in
table \ref{tab1}, it is clear that in order to achieve spectroscopic accuracy,
much larger-scale configuration interaction models than considered here are required.
This is  due primarily  to the very slow convergence of the external electron correlation
incorporated into the models. Due to the limitations of available computation resources,  even with access to world-leading computer facilities,
the 594-level approximation was used as a compromise
for the target and scattering approximations. It will be seen to
reproduce the main features in the experimental PI cross sections.

The atomic structure calculations were performed
with the GRASP code \cite{Dyall1989,Parpia2006,Grant2007} to provide target wavefunctions
for the various models of the residual ion  used in the close-coupling
PI cross section calculations.
All these calculations were performed in
 the Dirac Coulomb {\textit R}-matrix approximation
 using the parallel version of the DARC
 codes \cite{darc,McLaughlin2012a,McLaughlin2012b}.
 Fourteen continuum orbitals were employed to span the
photon energy range of interest and a boundary radius
of 29.76 atomic units was necessary to
 accommodate all the diffuse $n$=5 orbitals of the
 residual doubly ionized calcium ion.

PI cross section calculations
were performed for the $3s^2 3p^6 4s~^2{\textrm S}_{1/2}$
ground level  and the metastable
$3s^2 3p^6 3d~^2{\textrm D}_{3/2,5/2}$ levels of Ca$^+$. The outer-region electron-ion collision
 problem was solved (in the resonance region below and
 between all thresholds) using a suitably chosen fine
energy mesh of 8.0$\times$10$^{-7}$ Rydbergs ($\approx$ 10.9 $\mu$eV)
to resolve the fine resonance structure in the
 photoionisation cross sections.
The $jj$-coupled Hamiltonian diagonal matrices
were adjusted so that the theoretical term
energies matched the recommended experimental values of NIST \cite{NIST2016}.

\section{Results and Discussion}
\label{Sec:results}

\begin{figure}
\begin{center}
\includegraphics[width=11cm]{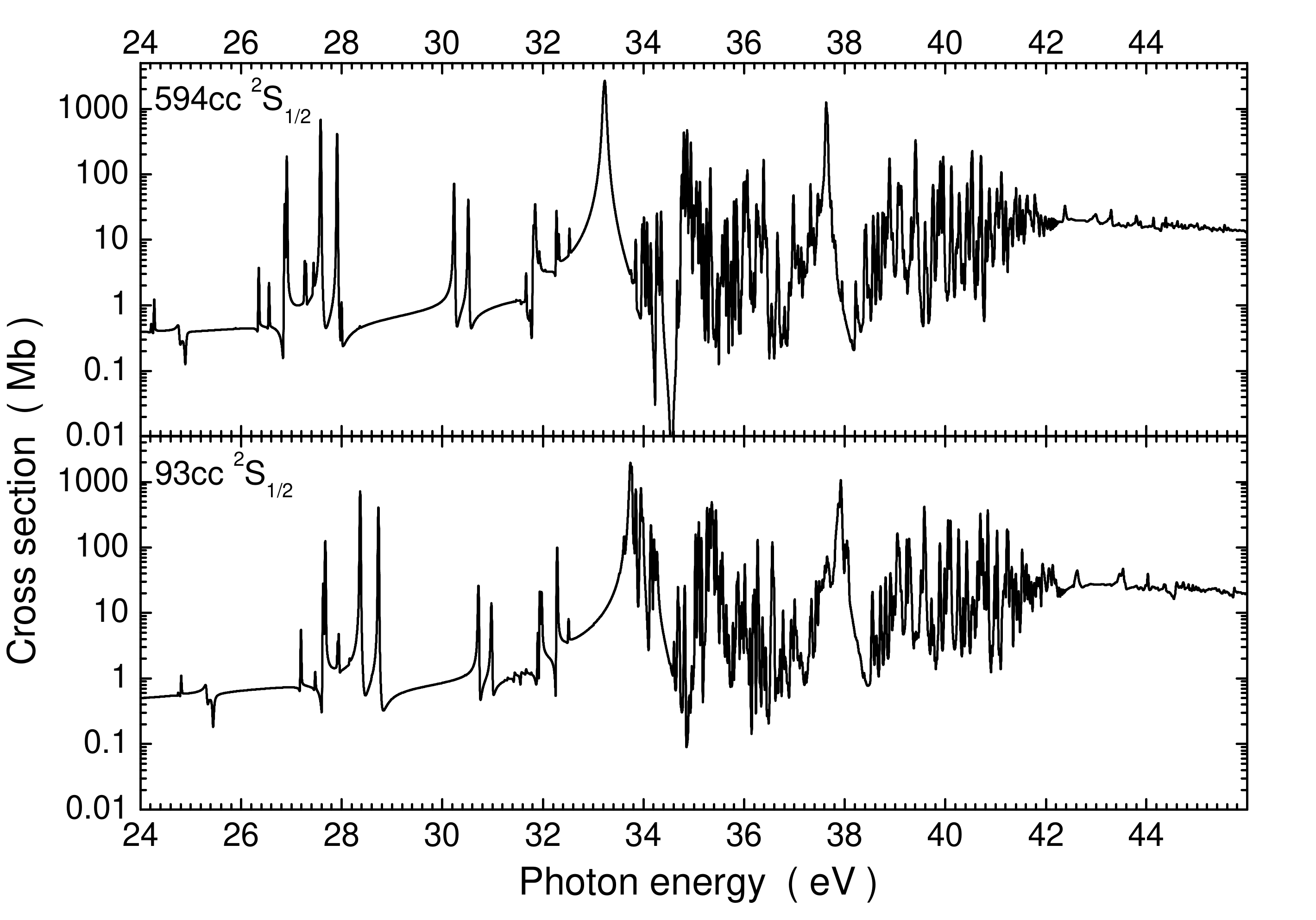}
\caption{\label{Fig:DARC2S1} Present DARC cross sections for PI of Ca$^+(3p^6 4s~^2{\textrm S}_{1/2})$ ions in their ground level. The upper panel shows the result of the 594-level close-coupling calculation (594cc), the lower panel that of the 93-level (93cc) approximation. The theoretical cross sections were convoluted with a Gaussian distribution of 17~meV FWHM corresponding to the photon energy spread in the present experimental overview spectrum. The displayed theoretical spectra are limited to the photon energy range where resonances occur. }
\end{center}
\end{figure}

\begin{figure}
\begin{center}
\includegraphics[width=11cm]{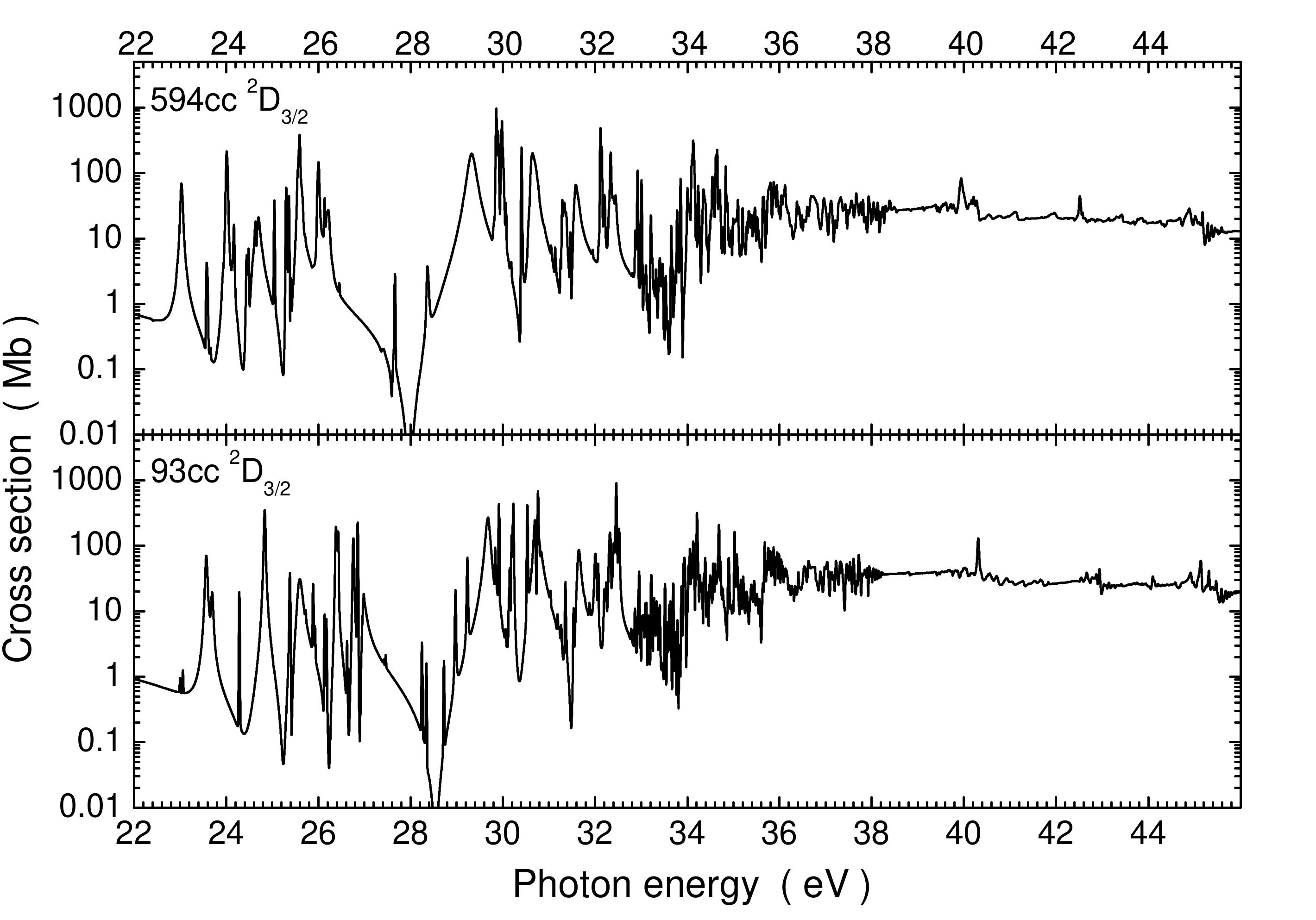}
\caption{\label{Fig:DARC2D3} Present DARC cross sections for PI of Ca$^+$ ions in the $3p^6 3d~^2{\textrm D}_{3/2}$ metastable level. For details see the caption of figure~\ref{Fig:DARC2S1}.}
\end{center}
\end{figure}

\begin{figure}
\begin{center}
\includegraphics[width=11cm]{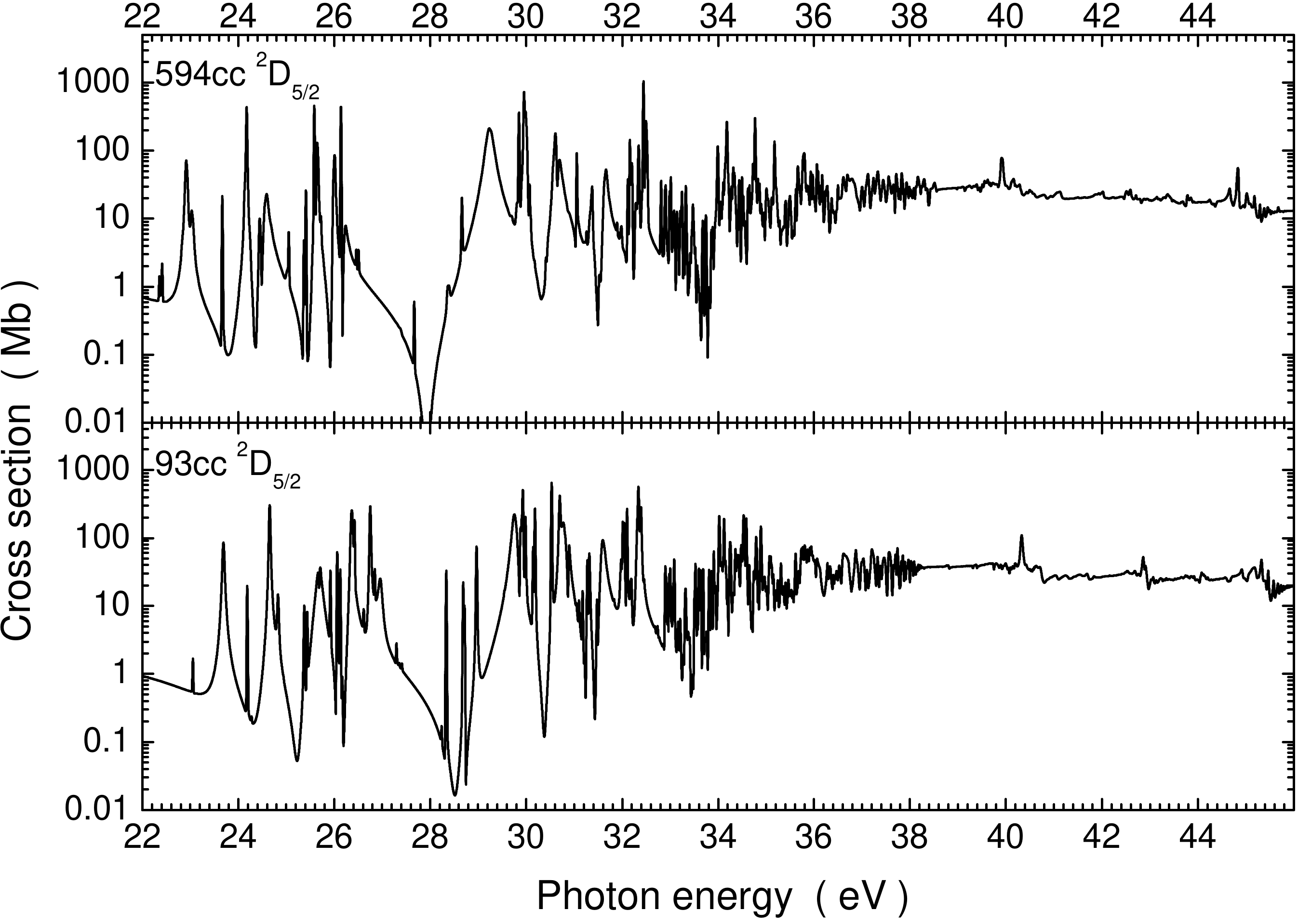}
\caption{\label{Fig:DARC2D5} Present DARC cross sections for PI of Ca$^+$ ions in the $(3p^6 3d~^2{\textrm D}_{5/2})$ metastable level. For details see the caption of figure~\ref{Fig:DARC2S1}.}
\end{center}
\end{figure}

The results of the present DARC PI cross-section calculations on the basis of the 93-level and 594-level models are presented in figure~\ref{Fig:DARC2S1} for Ca$^+(3p^6 4s~^2{\textrm S}_{1/2})$, in figure~\ref{Fig:DARC2D3} for Ca$^+(3p^6 3d~^2{\textrm D}_{3/2})$, and in figure~\ref{Fig:DARC2D5} for Ca$^+(3p^6 3d~^2{\textrm D}_{5/2})$. The sizes of the cross sections which were convoluted with Gaussians of 17~meV full width at half maximum (FWHM)  show energy-dependent variations of more than five orders of magnitude. Therefore, a logarithmic scale is chosen for the vertical axes. The cross sections obtained by the two different models are similar. Slight shifts in the energy scales, mainly towards lower energies for the more sophisticated calculations, are visible. The (non-resonant) cross sections for direct removal of an electron from Ca$^+$ at energies beyond approximately  42~eV are smaller in the 594cc approximation as compared to the 93cc model. Very broad strong resonances are found at about 33~eV in the ground-level Ca$^+$  PI spectrum and at about 29~eV in the PI spectra of metastable Ca$^+$ ions. The profiles of these ''giant'' resonances are jagged in the 93cc approximation and smooth in the 594cc calculations. The smooth profiles agree better with the experimental data as do the level energies of the broadest peaks obtained in the 594cc calculation (see below).

\begin{figure}
\begin{center}
\includegraphics[width=\columnwidth]{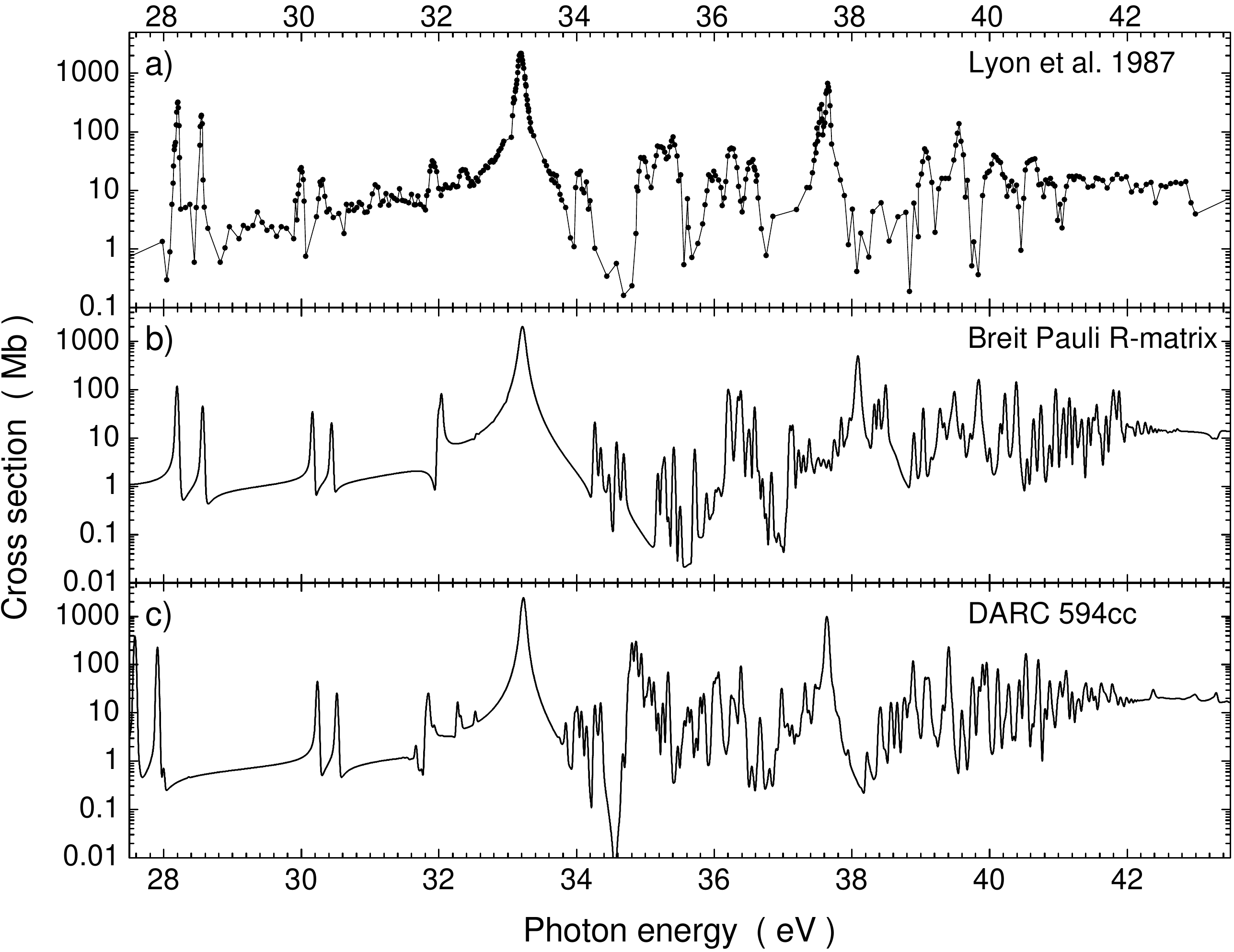}
\caption{\label{Fig:Lyoncomp}Comparison of the experimental results of Lyon~\etal~\cite{Lyon1987a} for the ground-level Ca$^+(3p^6 4s~^2{\textrm S}_{1/2})$ ion (panel a) with the 17-level Breit-Pauli {\textit R}-matrix calculation by Sossah~\etal~\cite{Sossah2012} (panel b) and with the present 594-level DARC  calculations (panel c). The theoretical spectra were convoluted by Gaussians of 32~meV FWHM in order to simulate the experimental resolution. Note the logarithmic cross-section scales. The experimental data points are connected by solid straight lines. }
\end{center}
\end{figure}

The PI calculations for ground-level Ca$^+(3p^6 4s~^2{\textrm S}_{1/2})$ ions can be compared with and benchmarked by the absolute measurements carried out by Lyon~\etal~\cite{Lyon1987a}. Their experimental PI spectrum was measured at a resolution of 32~meV. It is displayed in panel a) of figure~\ref{Fig:Lyoncomp}. For comparison, the theoretical results of the 17-level Breit-Pauli {\textit R}-matrix~\cite{Sossah2012} and the present 594-level DARC  calculations are shown in panels b) and c), respectively. The theoretical spectra were convoluted by Gaussians of 32~meV FWHM in order to simulate the experimental photon-energy resolution. At energies up to about 34~eV the main features in the experimental spectrum can be readily associated with corresponding structures in the theoretical spectra. At higher energies it is more difficult to find corresponding features in the experimental and theoretical spectra. Slightly below 38~eV there is another broad, strong peak in the measured cross section which is also visible in the theoretical spectra. It is reproduced better  by the present DARC calculation as compared to the Breit-Pauli result. The theoretical data reproduce the dominant features and overall spectral shape found in the experiment of Lyon~\etal reasonably well but deviate in the details of the smaller resonance contributions.

\begin{figure}
\begin{center}
\includegraphics[height=16.5cm]{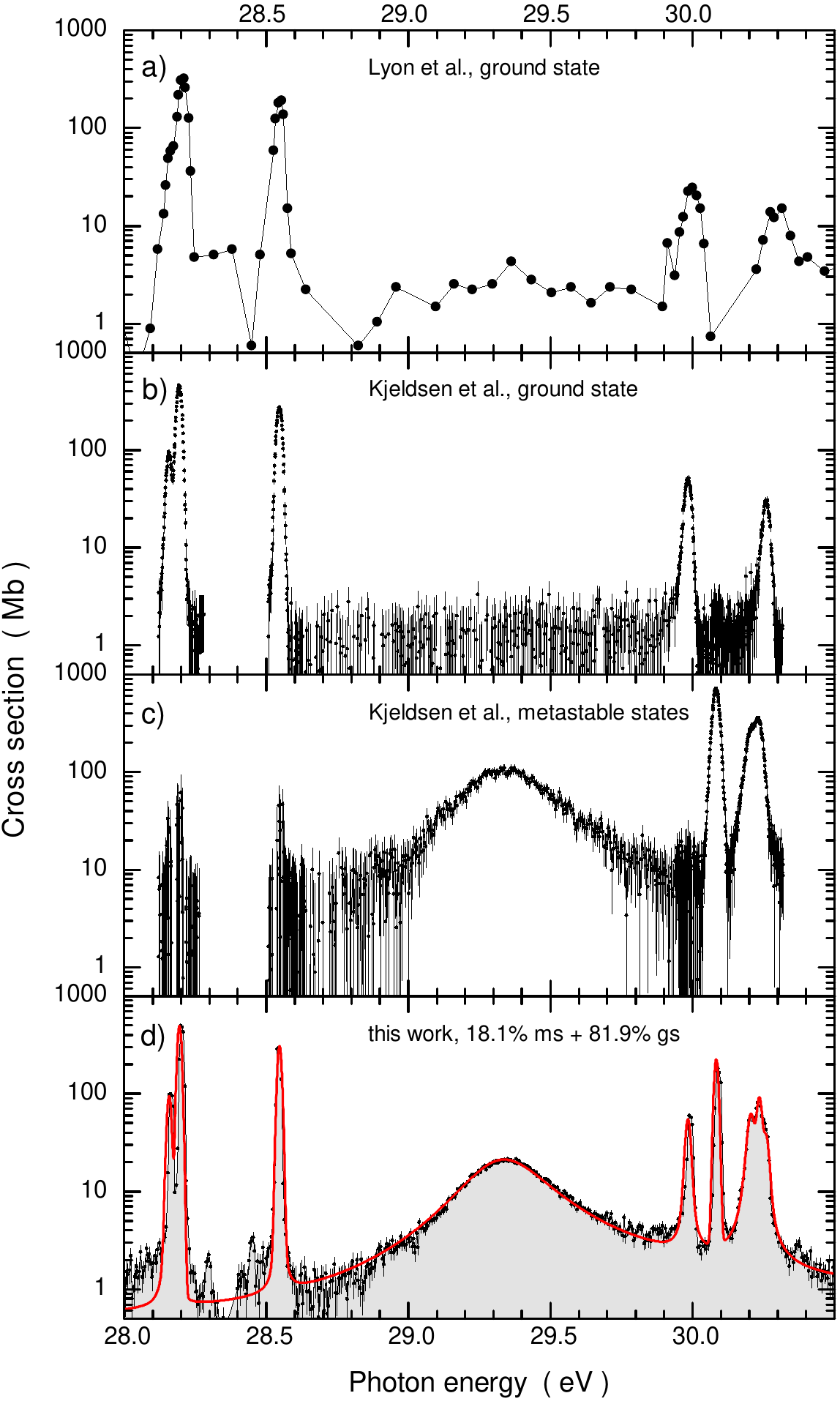}
\caption{\label{Fig:ALSnormalise}(Colour online) Cross sections for PI of Ca$^+$ ions in the narrow energy range investigated by Kjeldsen~\etal~\cite{Kjeldsen2002d}. The absolute cross sections for PI of ground-level Ca$^+(^2{\textrm S})$ measured by Lyon~\etal~\cite{Lyon1987a} (panel a) and by Kjeldsen~\etal~\cite{Kjeldsen2002d} (panel b) are compared to the absolute cross sections for PI of metastable Ca$^+(^2{\textrm D})$ derived by Kjeldsen~\etal~\cite{Kjeldsen2002d} (panel c). Panel d) displays the present normalised cross section for a mix of metastable and ground-level Ca$^+$ ions. The fraction  $m = 0.18 \pm 0.04$ of metastable ions in the parent ion beam was inferred for these measurements by comparison with the results shown in panels b) and c). The solid (red) line in panel d) shows the weighted sum of the cross sections from panels b) and c) with weights $(1-m)$ and $m$, respectively, multiplied by a factor 1.055 (see text) and taking the differences in photon energy resolution into account.
}
\end{center}
\end{figure}

\begin{figure*}
\begin{center}
\includegraphics[width=\columnwidth]{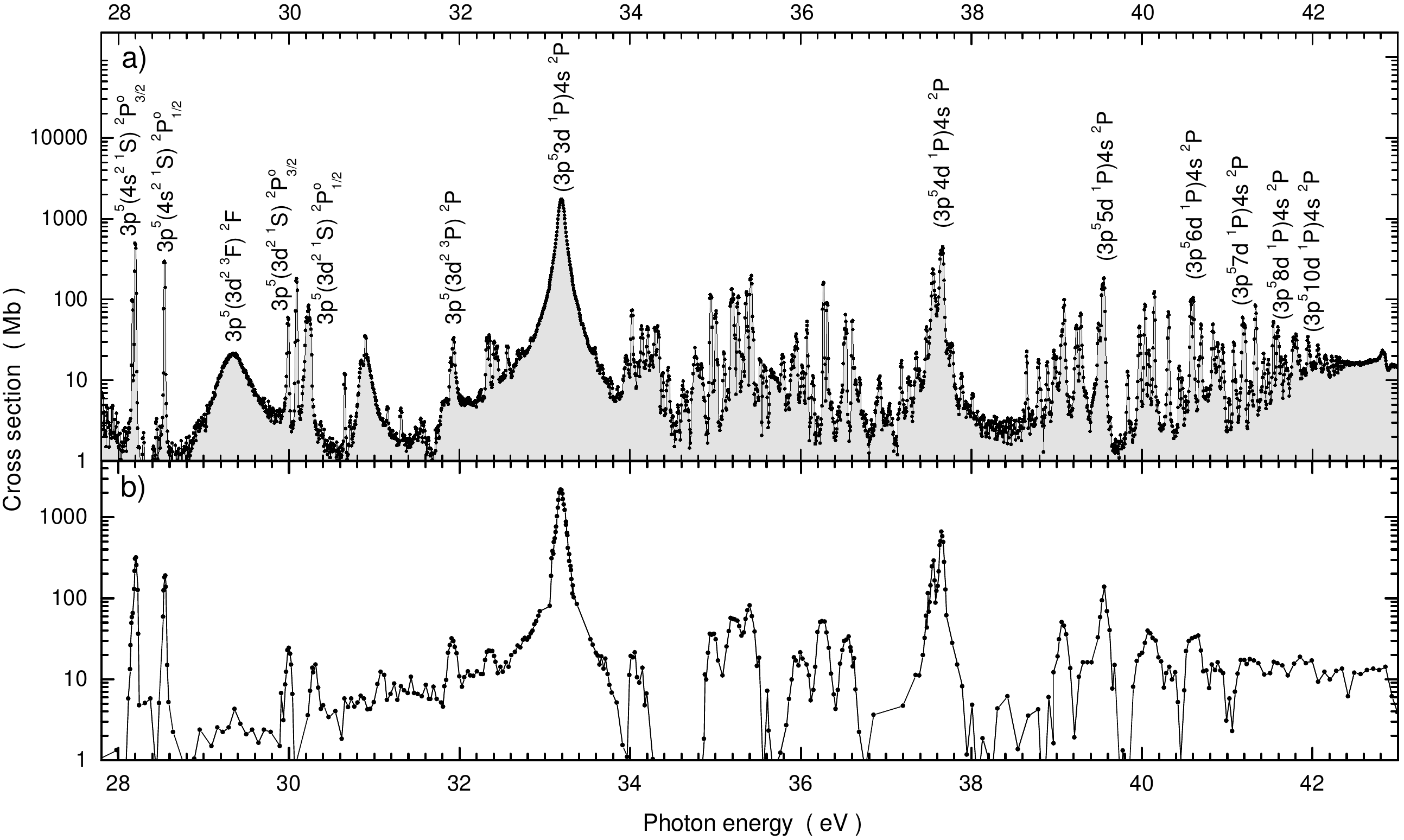}
\caption{\label{Fig:compALSLyon}Comparison of the present normalised cross sections for PI of a mixture of ground-level (82\%) and metastable (18\%) Ca$^+$ ions (panel a) with the absolute cross sections for PI of ground-level Ca$^+(^2{\textrm S})$ measured by Lyon~\etal~\cite{Lyon1987a} (panel b). The plot is restricted to the energy range of the data of Lyon~\etal$\!$. The bandwidths of photon energy distributions are 17~meV and 32~meV respectively. Spectral assignments are indicated for some of the most prominent resonances.
					}
\end{center}
\end{figure*}

Measurements by Kjeldsen~\etal~\cite{Kjeldsen2002d} on PI of ground-level and metastable Ca$^+$ in the energy range 28.0 to 30.5~eV are key to the normalization of the present experimental spectrum. Their absolute cross sections for PI of ground-level Ca$^+$ are shown in panel b) of figure~\ref{Fig:ALSnormalise} along with the data obtained in the pioneering experiment by Lyon~\etal~\cite{Lyon1987a} in the same energy range (see panel a) three decades ago. The superior resolution and statistical precision of the Kjeldsen~\etal data is obvious, emphasising advances in synchrotron radiation sources. A fit to the latter data yields a resolution of 21.2~meV compared to the 32~meV of the earlier measurement. The cross section derived by Kjeldsen~\etal for the metastable Ca$^+(3d~^2{\textrm D})$ parent-ion term is shown in panel c). A fit to these data suggests an energy resolution of about 28~meV.

The spectra in panels b) and c) were used to normalise the present relative cross sections for PI of a mixture of ground-level and metastable Ca$^+$ parent ions (panel d) as described already in section~\ref{sec:experiment}. The present data are superior to the results of Kjeldsen~\etal$\!$ with respect to resolution (17~meV) and statistical significance. The contributions of ground-level and metastable parent ions can be clearly and unambiguously distinguished at energies up to 30.13~eV. Only the multi-peak feature at photon energies between 30.14 and 30.30~eV is a mixture of contributions from the $^2{\textrm S}$ and $^2{\textrm D}$ initial terms and has therefore been excluded from the analysis. The comparison of the spectra in panels b), c) and d) provided the fraction $m = 0.18 \pm 0.04$ of metastable Ca$^+(^2{\textrm D})$ in the parent ion beam used in the present experiments. Since the cross sections obtained by Lyon~\etal for Ca$^+(^2{\textrm S})$ parent ions are about 11\% above the Kjeldsen~\etal result the present photoion yield data were normalised to the average of the two data sets for ground-level Ca$^+(^2{\textrm S})$ parent ions, i.e. they were multiplied by another factor of 1.055 beyond the correction by the factor $\alpha$ derived from the Kjeldsen~\etal data alone (see section~\ref{sec:experiment} and equation~\ref{Eq:apparentxsection}).

The consistency of the data analysis is demonstrated by comparing the present PI spectrum for a mixed beam of metastable and ground-level ions (data points in panel d; cross section $\sigma$) with the weighted sum (the solid (red) line in panel d) of the cross sections $\sigma_g$ (panel b) and $\sigma_m$ (panel c) obtained by Kjeldsen ~\etal~\cite{Kjeldsen2002d} with $ \sigma = 1.055 [m \sigma_m + (1-m) \sigma_g]$, $m=0.18$. For the comparison five and four Voigt profiles, respectively, on top of constant background cross sections  were  fitted to the spectra for ground-level and metastable Ca$^+$ measured by Kjeldsen~\etal$\!$. With the resonance parameters thus obtained, each of the two spectra, $\sigma_g$ and $\sigma_m$, could be represented by a smooth function and the resolution adjusted to the 17~meV of the present experiment. There is a slight shift in energy between the present spectrum and the data of Kjeldsen~\etal$\!$. The latter spectrum is shifted with respect to the present experiment by approximately 8~meV towards lower photon energies, well within the calibration uncertainty of $\pm 10$~meV estimated by Kjeldsen~\etal and only slightly outside the present range of uncertainty ($\pm 5$~meV).

\begin{figure*}
\begin{center}
\includegraphics[width=\columnwidth]{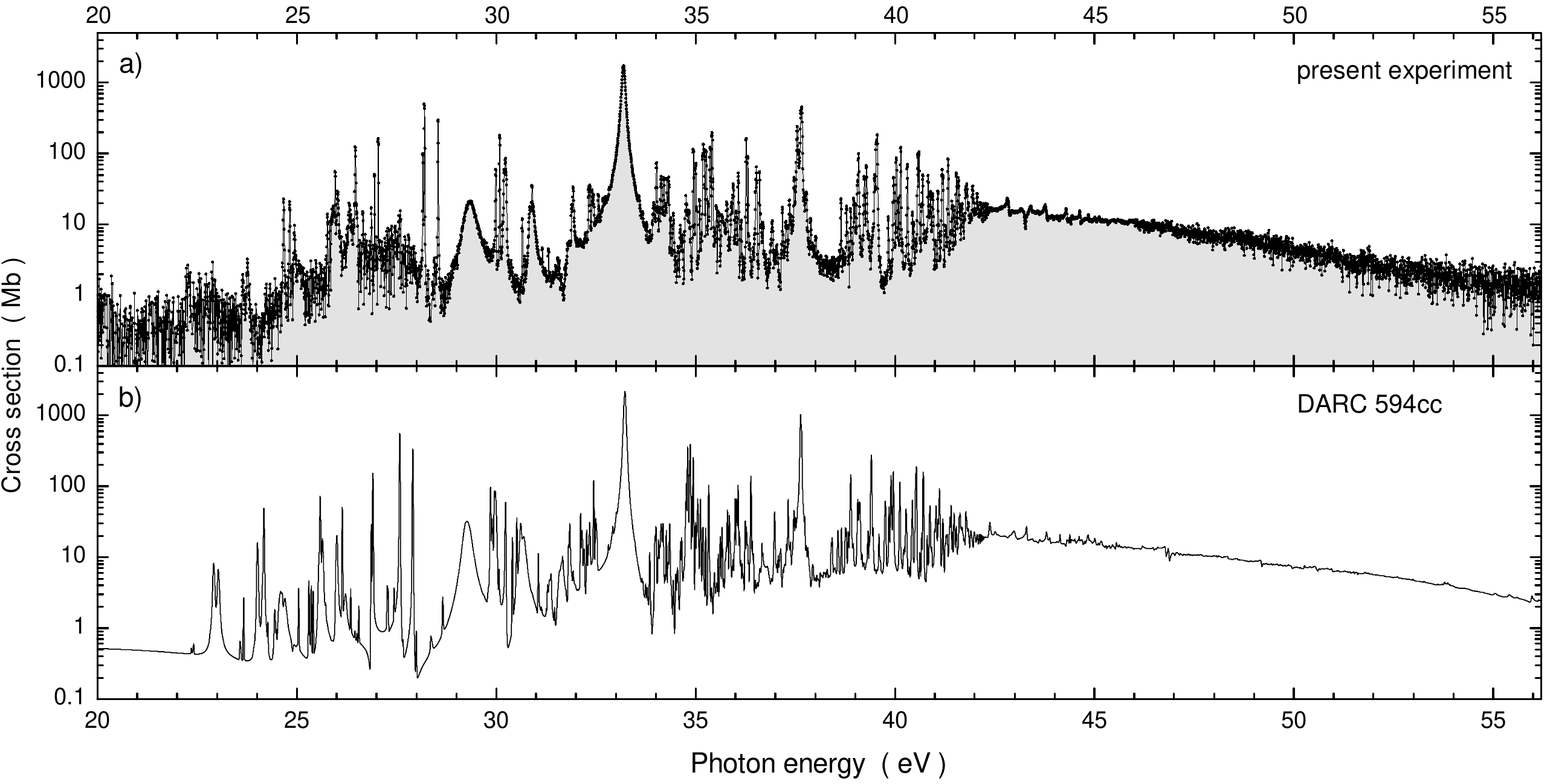}
\caption{\label{Fig:totoverviewcomp} Overview of the present experimental (panel a) and theoretical (panel b) results  obtained for PI of Ca$^+$ ions in the energy range 20 to 56~eV. The experimental data are the same as those displayed in figure~\ref{Fig:compALSLyon} but now the results for the full energy range investigated in the present measurements are shown. For the comparison with the present 594-level DARC calculations the specific experimental conditions had to be accounted for. The experimental data were taken at a photon-energy resolution of 17~meV. Hence, the theoretical results were convoluted with a Gaussian of 17~meV FWHM, assuming that the spectral distribution of the photon beam can be approximated by a Gaussian function. The ion beam used in the experiment consisted of a fraction of 82\% of Ca$^+$ ions in the $^2{\textrm S}_{1/2}$ ground level and 18\% of Ca$^+$ ions in the $^2{\textrm D}$ metastable term with fine-structure levels $^2{\textrm D}_{3/2}$ and $^2{\textrm D}_{5/2}$. Assuming that these fine-structure levels were populated according to their statistical weights, an appropriately weighted sum of the theoretical results shown in figures~\ref{Fig:DARC2S1}, \ref{Fig:DARC2D3} and  \ref{Fig:DARC2D5} was calculated to model the experimental results and is shown in panel b.
}
\end{center}
\end{figure*}

While the data taken by Kjeldsen~\etal~\cite{Kjeldsen2002d} on PI of Ca$^+$ are restricted to the energy range 28.0 to 30.5~eV the pioneering measurement carried out by Lyon~\etal~\cite{Lyon1987a} covers a range from 27~eV to 43~eV. Figure~\ref{Fig:compALSLyon} compares the present normalised PI spectrum of Ca$^+$ with the absolute cross sections for PI of ground-level Ca$^+([{\textrm Ar}]4s~^2{\textrm S})$ measured by Lyon~\etal$\!$. Since the parent ion beam used in the present experiment consisted of two components,  Ca$^+([{\textrm Ar}]4s~^2{\textrm S})$ and Ca$^+([{\textrm Ar}]3d~^2{\textrm D})$, the spectrum displayed in panel a) shows peak features arising from the metastable component (fractional abundance 18\%) that are not present in the data in panel b) obtained with a pure ground-level Ca$^+$ ion beam. One of the more prominent additional peaks with a large width seen in the present PI spectrum is associated with the $3p^5 (3d^2~^3{\textrm F})~^2{\textrm F}$ term near 29.3~eV which is populated by excitation of metastable Ca$^+(3d~^2{\textrm D})$ ions.

Very strong, broad resonances arising from photoexcitation of ground-level Ca$^+(^2{\textrm S})$ ions are known from the early experimental work of Lyon~\etal and since the theoretical treatment by Miecznik~\etal~\cite{Miecznik1990}. These ''giant'' resonances are associated with Ca$^+((3p^5 nd~^1{\textrm P}) 4s~^2{\textrm P})$  autoionizing terms. The strongest of these $3p \to nd$  resonances with $n=3$ is near 33~eV. Its peak is at 2190~Mb in the Lyon~\etal data and at 1731~Mb in the present spectrum where the ground-level fraction is only 82\%. When correcting for this fractional abundance, the present experiment yields a peak cross section of 2114~Mb, only 3.5\% below the Lyon~\etal measurement, and well within their systematic experimental uncertainties. Higher members of the $(3p^5 nd~^1{\textrm P}) 4s~^2{\textrm P}$ Rydberg sequence are seen in the Lyon~\etal spectrum ($n=4,5$) and in the present spectrum shown in panel a) with $n$ up to 12. A careful analysis and higher-resolution measurements reveal Rydberg contributions up to $n=22$ (see below). The peak assignments provided in figure~\ref{Fig:compALSLyon} are discussed in more detail below in the context of observed level energies and peak widths.

Figure~\ref{Fig:totoverviewcomp} compares the present experimental and theoretical results obtained in the photon energy range 20 to 56~eV. Since the parent ion beam in the experiment was a mixture of Ca$^+(3p^6 4s ~^2{\textrm S}_{1/2})$ ions in their ground level and of Ca$^+(3p^6 3d ~^2{\textrm D}_{3/2,5/2})$ in their metastable levels, a weighted sum of the theoretical cross sections for each of these levels had to be constructed to model the experiment. The fractional abundances of ground-level and metastable-term ions have been inferred to be 82\% and 18\%, respectively (see section~\ref{sec:experiment}). For the relative population of the fine-structure components $^2{\textrm D}_{3/2}$ and $^2{\textrm D}_{5/2}$ of the metastable $^2{\textrm D}$ term statistical weights of 4/10 and 6/10, respectively, are a reasonable assumption (see for example~\cite{Mueller2010b}). The cross sections $\sigma(^2{\textrm S}_{1/2})$, $\sigma(^2{\textrm D}_{3/2})$, and $\sigma(^2{\textrm D}_{5/2})$ for the participating initial levels are known from the present 594-level DARC calculations (see figures~\ref{Fig:DARC2S1}, \ref{Fig:DARC2D3} and  \ref{Fig:DARC2D5}). Thus, a model cross section $\sigma$ representing the experimental spectrum was calculated as $\sigma = 0.819 \sigma(^2{\textrm S}_{1/2}) + 0.18 [ 0.4 \sigma(^2{\textrm D}_{3/2}) + 0.6 \sigma(^2{\textrm D}_{5/2})]$. It was convoluted with a Gaussian distribution of 17~meV FWHM to represent the experimental spectral distribution. The result is the solid line in panel b) of figure~\ref{Fig:totoverviewcomp}.
\begin{figure*}
\begin{center}
\includegraphics[width=\columnwidth]{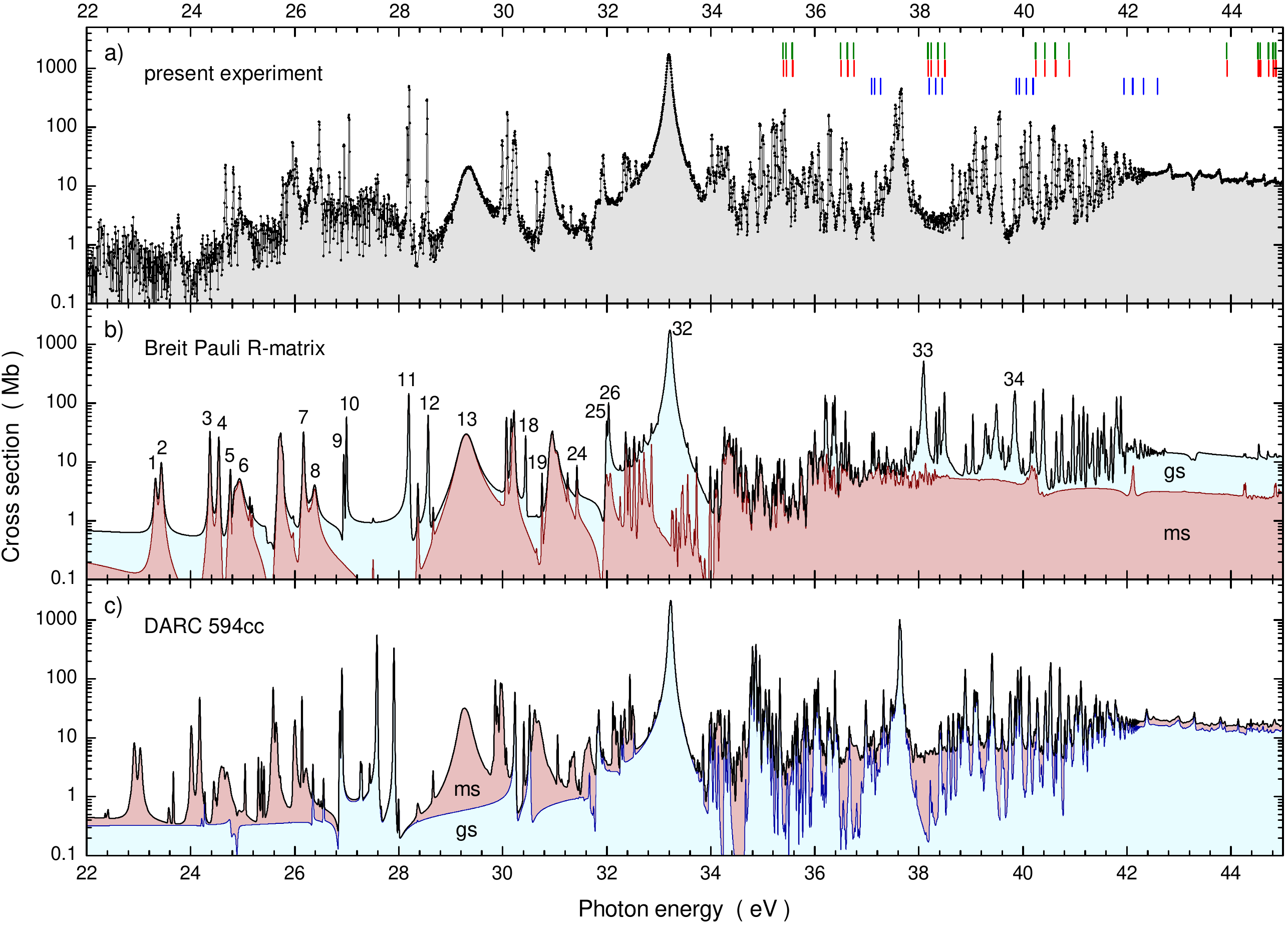}
\caption{\label{Fig:overviewcompnew}(Colour online) Region where resonances occur in the valence-shell PI of Ca$^+$ ions. Panel a) shows the present experimental PI spectrum. Three rows of vertical bars indicate thresholds for direct photoionisation of Ca$^+(3p^6 4s ~^2{\textrm S}_{1/2})$ ions (lowest row of (blue) vertical bars), of  Ca$^+(3p^6 3d ~^2{\textrm D}_{3/2})$ (middle row of light (red) vertical bars) and Ca$^+(3p^6 3d ~^2{\textrm D}_{5/2})$ (upper row of dark (green) vertical bars). The final states are $3p^5 3d$ and $3p^5 4s$ excited levels of the Ca$^{2+}$ ion as presented in table~\ref{tab1}. Panel b) displays the model cross section derived from the Breit-Pauli {\textit R}-matrix calculations performed by Sossah~\etal~\cite{Sossah2012} similar to the one presented in panel b) of figure~\ref{Fig:totoverviewcomp} obtained from the present DARC calculations which are shown again in panel c). In both theory spectra the contributions from the Ca$^+(3p^6 4s ~^2{\textrm S}_{1/2})$ ground level and the Ca$^+(3p^6 3d ~^2{\textrm D})$ metastable term are separately specified. In panel b) the ground-level contribution of the cross section (light (blue) shading) is shown on top of the metastable-term contribution of the cross section (dark (red) shading). In panel c) the metastable-term contribution of the cross section (dark (red) shading) is shown on top of the ground-level contribution of the cross section (light (blue) shading). The numbers in panel b) correspond to level identifications (see table~\ref{Tab:identification}) by Sossah~\etal~\cite{Sossah2012}, except for numbers 33 and 34.
}
\end{center}
\end{figure*}

The experimental data show considerable statistical fluctuations in the regions of the lowest and the highest photon energies. This is due to the small cross sections and the resulting small signal rates. Moreover, the photon flux from the grating used in this experiment is low at energies below about 25~eV and above 45~eV. Nevertheless, small resonances with magnitudes of only a few Mb could be still investigated even in the unfavorable photon-energy regions. Larger resonances were observed with excellent statistical quality. The comparison with the DARC calculations shows good overall agreement of the experimental and theoretical cross sections. In particular, the magnitudes, shapes and energies of the most prominent features are well reproduced by theory.

An additional check on the theoretical data  is a comparison of the integrated oscillator strength $f$ with experiment.
The integrated oscillator strength $f$  of the PI spectra was calculated using \cite{Fano1968},
\begin{equation}
\label{Eq:strength}
f = 9.1075 \times 10^{-3}\,(\textrm{Mb\,eV})^{-1} \int_{20~eV}^{55~eV} \sigma (E) dE.
\end{equation}
The total oscillator strength  from the present ALS measurements is 5.2 $\pm$ 1.6.
 A similar procedure for the 594-level DARC theoretical  cross-section model (from appropriately weighted initial states) gives a
value of 5.69.
The Breit-Pauli calculations of Sossah et al \cite{Sossah2012} yield 4.72.
Both theory values are well within the experimental uncertainty. The oscillator strengths found in the energy range between 20 and 55~eV  by theory and experiment account for most of the total oscillator strength ( 6 for 6 electrons) associated with the $3p$ subshell. No significant resonance contributions arising from excitation of a $3p$ electron are expected outside the present energy range.

For a more detailed comparison of theory and experiment figure~\ref{Fig:overviewcompnew} focuses on the energy region where resonances are observed. Different from figure~\ref{Fig:totoverviewcomp} it includes the results obtained by Sossah~\etal and shows the contributions of the metastable and ground-level parent ions separately. The experimental data in panel a) are the same as those displayed in panel a) of figure~\ref{Fig:totoverviewcomp}. Resonance features in the investigated energy range are primarily due to promotions of $3p$ electrons. Excitation of a $3s$ electron requires energies of about 50~eV or higher.  PI thresholds for populating individual final excited levels of Ca$^{2+}$ from Ca$^{+}$ are indicated by three rows of solid vertical bars in the upper right region of panel a). They represent the onsets of processes such as $\gamma + {\textrm {Ca}}^+(3p^6 4s ~^2{\textrm S}_{1/2}) \to e^- + {\textrm {Ca}}^{2+}(3p^5 4s ~^1{\textrm P}_1)$.  The lowest row is for direct PI from the $^2{\textrm S}_{1/2}$ ground level, the middle and upper rows of vertical bars are for the $^2{\textrm D}_{3/2}$ and $^2{\textrm D}_{5/2}$ metastable initial levels, respectively, which do not differ much from one another. The threshold energies are given by the minimum ionisation energy of the initial Ca$^+$ level plus the excitation energy from the ground state of Ca$^{2+}$ to the given excited level. The first sixteen of these levels are discussed in table~\ref{tab1}. All energies were taken from the NIST tabulations~\cite{NIST2016}.

All excited levels of Ca$^{2+}$ that can be reached from one of the Ca$^+$ ground and metastable initial levels are associated with Rydberg sequences of highly excited, autoionizing levels in Ca$^+$. Given the number of PI thresholds in the photon energy range of interest it is understandable that the PI spectrum is very complex, especially near the series limits. Assignment of levels to the observed peak features in the energy range above 34~eV is difficult. Identification was possible mainly at energies lower than 34~eV. Peaks identified by Sossah~\etal~\cite{Sossah2012} are marked by numbers in panel b) of figure~\ref{Fig:overviewcompnew} which shows the results of their Breit-Pauli {\textit R}-matrix calculations combined in a manner identical to the treatment of the present DARC calculations for modeling the present experiment. Not all of the resonances are marked in panel b) because of the limited space available. Complete information is provided in the context of the subsequent figures, tables and the associated text. Panel c) displays the results of the present 594-level DARC calculations modeling the experimental spectrum as described in the context of figure~\ref{Fig:totoverviewcomp}.

The individual contributions to the model cross sections arising from the $^2{\textrm S}_{1/2}$ ground level and the $^2{\textrm D}$ metastable excited term are specified in the results of the theoretical work shown in figure~\ref{Fig:overviewcompnew}. In panels b) and c)  the light (blue) shaded areas represent the ground-level contributions to the model spectra, the darker-shaded areas represent the metastable-term contributions. In panel b) the ground-level contribution is added to the metastable-term contribution while in panel c) the metastable-term contribution is added to the ground-level contribution. This makes a difference in the appearance of the relative contributions because the logarithmic scale emphasizes small cross sections. The different presentations of the two theory spectra, which are quite similar to one another, make this effect of the nonlinear scale obvious and help assess the magnitudes of the individual contributions. This comparison clarifies which peak features arise from the metastable-term and which from the ground level.

The following figures display enlargements of smaller energy regions of the experimental spectrum on linear scales so that fine details in the spectra are evident.
\begin{figure}
\begin{center}
\includegraphics[width=\columnwidth]{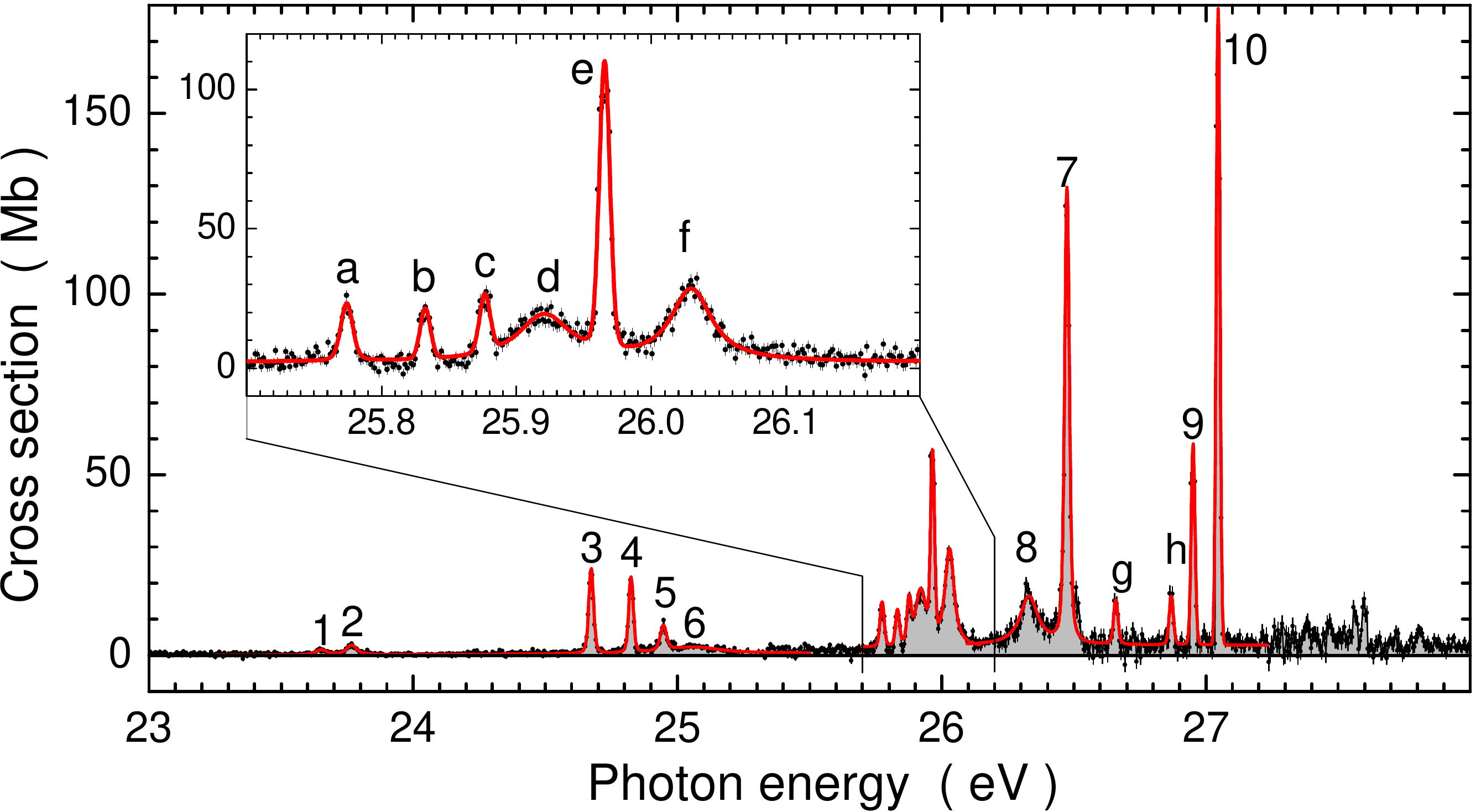}
\caption{\label{Fig:Eph23to28}(Colour online) Present experimental spectrum measured at 17~meV resolution in the energy region 23 to 28~eV (note the linear scale). The solid (red) lines are fits to the experimental data with six Voigt profiles each. Resonances that can be associated with level identifications provided by Sossah~\etal~\cite{Sossah2012} are labeled by numbers, additional features where the identification is not unique are labeled by letters. The inset displays the results of a short-range energy scan with a resolution of 8.8~meV. The solid (red) line in the inset is a separate fit to the experimental data using six Voigt profiles and considering the increase in resolution. Level assignments and  resonance parameters extracted from the fits are provided in tables~\ref{Tab:identification} and \ref{Tab:letters}.
}
\end{center}
\end{figure}
Figure~\ref{Fig:Eph23to28} shows the present experimental spectrum in the energy range 23 to 28~eV. Peaks labeled by numbers are associated with level identifications  provided by Sossah~\etal~\cite{Sossah2012}. The assignments are made on the basis of similarities of the present experimental results with the calculated Breit-Pauli model spectrum shown in figure~\ref{Fig:overviewcompnew}. Peaks labeled by letters could not be unambiguously identified. The complete set of cross-section features displayed in this figure  has been fitted with eighteen Voigt profiles shown as a solid (red) line. The narrow energy region 25.7 to 26.2~eV was scanned with a resolution of 8.8~meV at 1.5~meV step width. The resulting high-resolution spectrum is shown in the inset.  The numbered peaks are identified in table~\ref{Tab:identification} where the peak energies,
the Lorentzian peak widths inherent in the Voigt profiles, and the resonance strengths are also provided. These results are compared in the same table with the findings of Sossah~\etal in order to provide an assessment of the quality of their peak assignments. The fitting parameters  for the peaks labeled by letters are presented in table~\ref{Tab:letters}.

\begin{figure*}
\begin{center}
\includegraphics[width=\columnwidth]{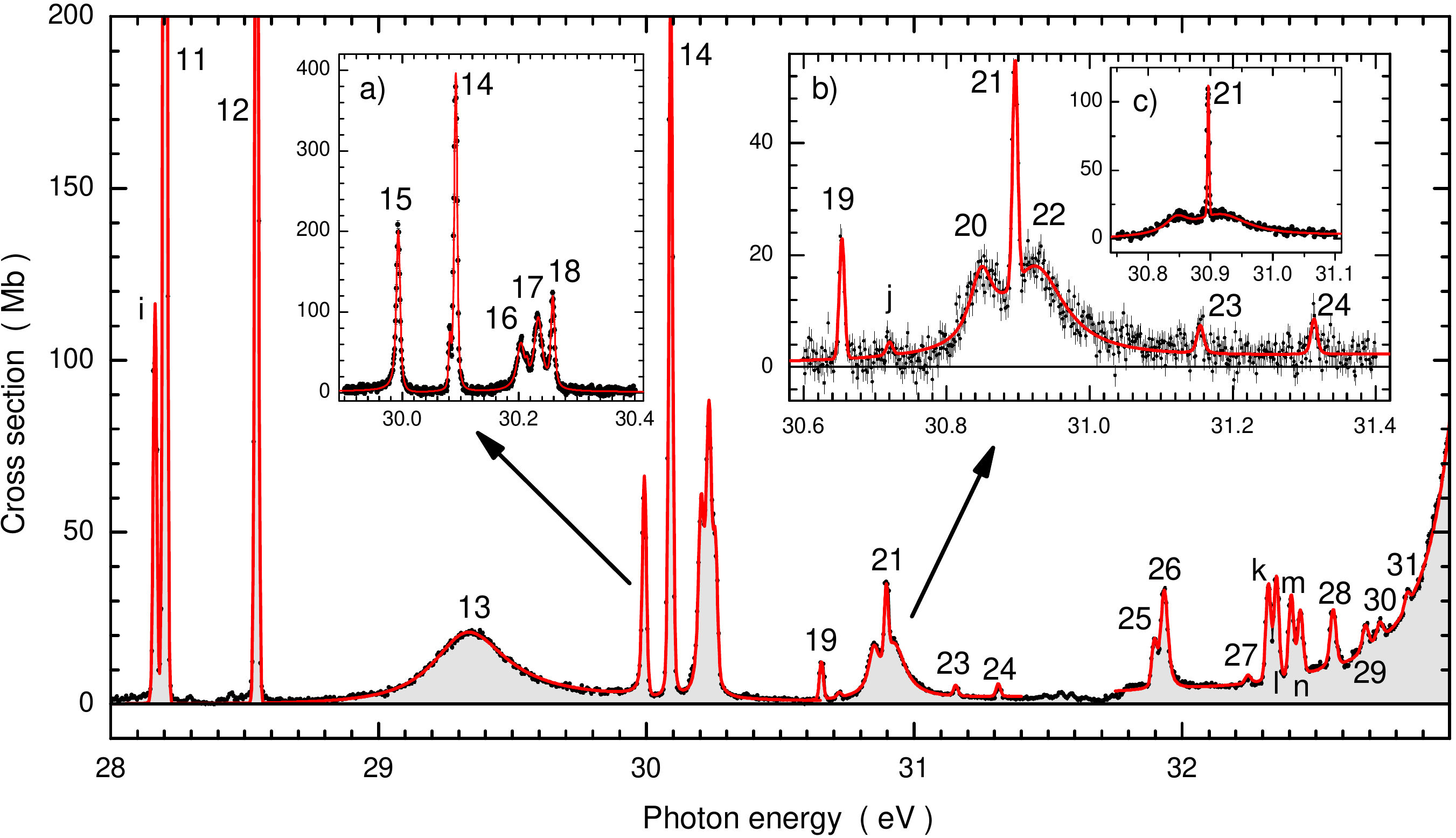}
\caption{\label{Fig:ms2FRegion}(Colour online)
 Present experimental spectrum measured at 17~meV resolution in the energy region 28 to 33~eV. In this photon energy range the strong $(3p^5 3d^2~^3{\textrm F})~^2{\textrm F^o}$ term can be excited from the metastable $^2{\textrm D}$ term of the Ca$^+$ ion. The solid (red) lines are fits to the experimental data. For the range 28 to 30.5~eV ten Voigt profiles and a Fano profile convoluted with a Gaussian~\cite{Schippers2011} (for the very broad $^2{\textrm F^o}$ resonance) were invoked. For energies beyond 31.8~eV a broader range was fitted with altogether twelve resonances (see figure~\ref{Fig:gs2Pregion}). Resonances that can be associated with level identifications provided by Sossah~\etal~\cite{Sossah2012} are labeled by numbers, additional features where the identification is not unique are labeled by letters. The left inset (panel a) displays the results of an energy scan from 29.9 to 30.4~eV with a resolution of 4.6~meV. The solid (red) line in this inset is a fit to the experimental data using seven Voigt profiles considering the increased resolution. The right inset (panel b) shows the results of an  energy scan from 30.6 to 31.4~eV with a resolution of 8.8~meV. The solid (red) line in this inset is a fit to the experimental data also using seven Voigt profiles and the specific resolution has been extracted from the fit. The inset (panel c) inside the right inset (panel b) displays an energy scan from 30.75 to 31.1~eV. The solid (red) line in this inset is a fit to the experimental data using three Voigt profiles. It yielded an experimental resolution of 3.3~meV.  Level assignments and extracted resonance parameters are provided in tables~\ref{Tab:identification} and \ref{Tab:letters}.
}
\end{center}
\end{figure*}

\begin{figure}
\begin{center}
\includegraphics[width=\columnwidth]{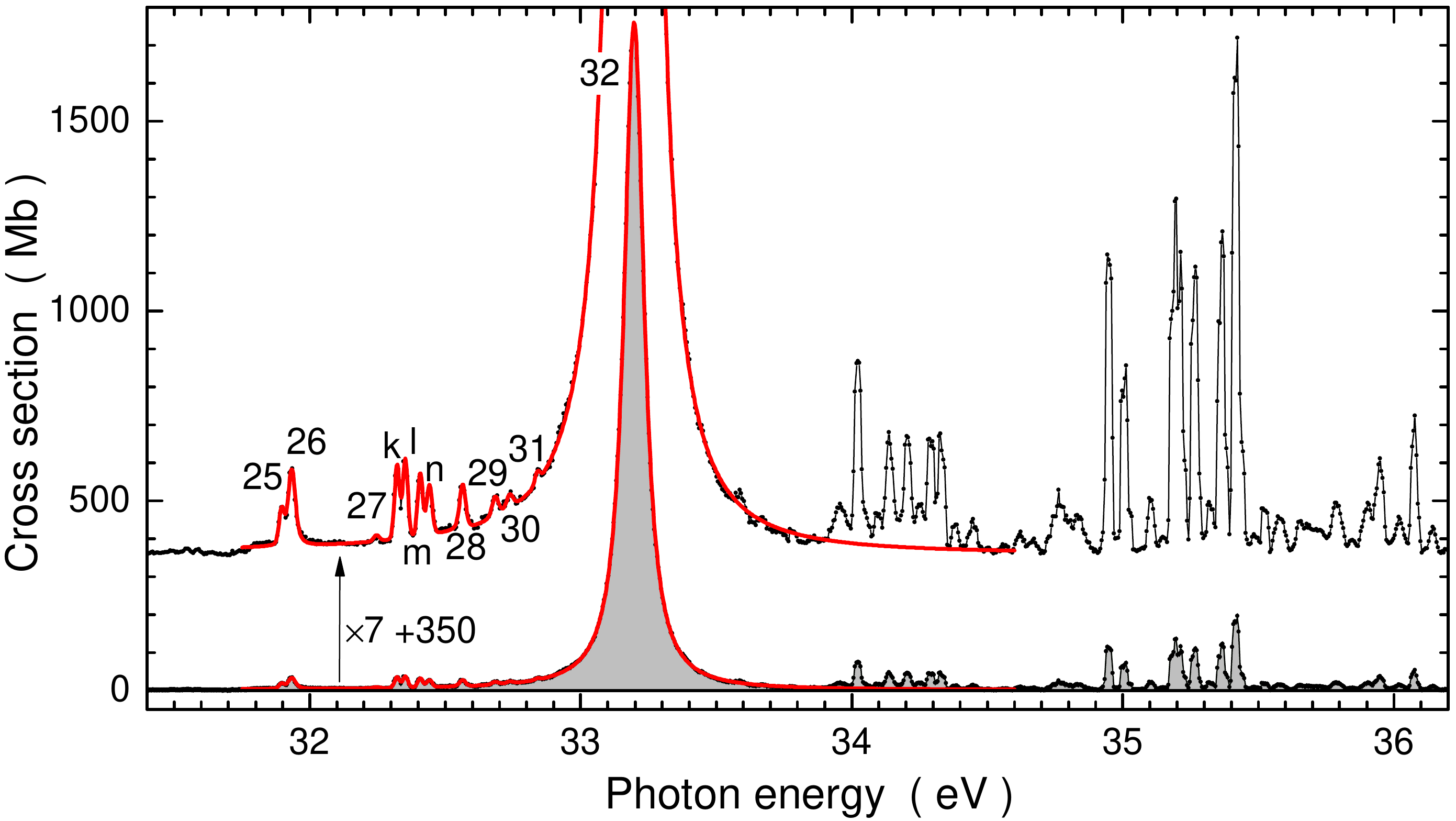}
\caption{\label{Fig:gs2Pregion}(Colour online)
 Present experimental spectrum measured at 17~meV resolution in the energy region 31.4 to 36.2~eV. In this photon energy range the strong $(3p^5 3d~^1{\textrm P})4s~^2{\textrm P}^o$ term can be excited from the  $^2{\textrm S}_{1/2}$ ground level of the Ca$^+$ ion. The solid (red) lines are fits to the experimental data. Resonances that can be associated with level identifications provided by Sossah~\etal~\cite{Sossah2012} are labeled by numbers, and additional features where the identification is not unique are labeled by letters. For emphasizing the smaller resonances, the spectrum has been multiplied by  a factor of 7 and then displayed again with an offset of 350~Mb. The solid (red) line is a fit to the experimental data using eleven Voigt profiles and one Fano profile convoluted with a Gaussian~\cite{Schippers2011}  for the ''giant'' resonance. The fit curve is also shown again with a multiplication factor of seven and an offset of 350~Mb.  Level assignments and extracted resonance parameters are provided in tables~\ref{Tab:identification} and \ref{Tab:letters}. No attempt was undertaken to identify all of the numerous additional resonances occurring at energies beyond 33.5~eV.}
 \end{center}
\end{figure}

Figure~\ref{Fig:ms2FRegion} shows the present experimental spectrum in the energy range 28 to 33~eV. As in the preceding figure, peaks are labeled by numbers and letters.  The complete set of cross section features displayed in this figure  (and partly in figure~\ref{Fig:gs2Pregion}) has been fitted with altogether twenty-eight Voigt profiles and, for the broadest peaks, two Fano profiles~\cite{Fano1968} convoluted with a Gaussian distribution function~\cite{Schippers2011}. The fit curve is shown as a solid (red) line. The narrow energy region 29.9 to 30.4~eV was scanned with a resolution of 4.6~meV at 0.8~meV step width. The resulting high-resolution spectrum is shown in the left inset.

The energy range around 30.9~eV shows an intriguing resonance feature which suggests interference of a broad and a narrow resonance of the same symmetry. In order to get additional information, the energy region 30.6 to 31.4~eV was scanned at 8.8~meV resolution with a step width of 1.5~meV. At this resolution the interference hypothesis is not compelling. A fit with three Voigt profiles for the central multi-peak feature gives a satisfactory result. This is even more so in the measurement with a further enhanced resolution of 3.3~meV at a step width of 0.5~meV. Clearly, three Voigt profiles give a good fit of the measured multi-peak feature. It is interesting to note, though, that the theoretical calculations of both the present 594-level DARC and particularly the previous Breit-Pauli {\textit R}-matrix models show clear signatures of interference of peak numbers 15 and 17 which arise solely from the metastable component of the parent ion beam. As for the preceding figure, the peak labels are explained in tables~\ref{Tab:identification} and \ref{Tab:letters}, respectively.

The strong increase of the PI cross section seen at energies beyond 32~eV in figure~\ref{Fig:ms2FRegion} is the low-energy tail of the well known ''giant resonance'' that had been observed previously by Lyon~\etal$\!$. It is associated with the autoionizing $(3p^5 3d~^1{\textrm P})4s~^2{\textrm P}^o$ term that is excited from the  $^2{\textrm S}_{1/2}$ ground level. This resonance dominates the whole spectrum as figure~\ref{Fig:gs2Pregion} clearly demonstrates. In order to resolve the smaller peaks which still have tens of Mb of cross section, the measured spectrum is multiplied by a factor 7 and shown again in figure~\ref{Fig:gs2Pregion} with an offset of 350~Mb. At energies beyond 33.5~eV, i. e., beyond the giant resonance the spectrum becomes very complex with many overlapping, unresolved resonances. Therefore, peak identification is restricted in the following to some of the more outstanding higher-energy features. The energy region below 33.5~eV  is still sufficiently simple that a good fraction of the observed peaks can be identified. The first half of the spectrum displayed in figure~\ref{Fig:gs2Pregion} has been fitted with eleven Voigt profiles plus a Fano profile convoluted with a Gaussian~\cite{Schippers2011} representing the ''giant resonance''. Their energies and widths are provided in tables~\ref{Tab:identification} and \ref{Tab:letters}.

While the fits give good representations of the experimental spectrum, one has to bear in mind that the broad peaks labeled with numbers 13 and 32 are not isolated single resonances. They are rather associated with unresolved $(3p^5 3d^2~^3{\textrm F})~^2{\textrm F^o}_{5/2,7/2}$ (peak 13) and $(3p^5 3d~^1{\textrm P})4s~^2{\textrm P}^o_{1/2,3/2}$ (peak 32) fine-structure levels. Moreover, resonance 13 can be excited from both metastable levels, Ca$^+(3p^6 3d ~^2{\textrm D}_{3/2})$ and Ca$^+(3p^6 3d ~^2{\textrm D}_{5/2})$. Thus, the two broad peaks are mixtures of several unresolved individual resonances and hence fitting them with a single Gaussian convoluted Fano profile~\cite{Schippers2011} each provides only an idea of the natural widths and the Fano asymmetry parameters $q$ of the individual resonances. This is particularly true for the $^2{\textrm F^o}$ resonance (13) where theory predicts the individual components to be up to 90~meV apart. The situation is better in the case of peak 32 which is populated from a single level, the $^2{\textrm S}_{1/2}$ ground level of the Ca$^+$ ion. The fine-structure splitting is of the order of only 10~meV which is small compared to the Lorentzian width of $\Gamma = 83 \pm 1$~meV found by fitting the experiment. The error bar of 1~meV is only the statistical uncertainty resulting from the fit of two fine-structure resonances with one Gaussian convoluted Fano profile~\cite{Schippers2011}. The Fano asymmetry parameters found for peaks numbered 13 and 32 are q=-19.8 and q=-174.7, respectively. Because of the blending of several resonances these numbers are only indicative for the presence of interference between direct and resonant PI channels associated with the experimental peak structures.

As mentioned above, the numbers and letters provided as labels in the preceding figures are explained in tables~\ref{Tab:identification} and \ref{Tab:letters}, respectively. From the fits to the peaks in those figures the resonance parameters were extracted. The peak areas are directly related to the oscillator strengths for photon absorption populating the associated excited levels. For an isolated resonance   the integration in equation~\ref{Eq:strength} has to be performed over the complete energy range to obtain the peak area from which the oscillator strength follows. Resonance energies, Lorentzian widths and resonance strengths found for the labeled peaks in figures~\ref{Fig:Eph23to28}, \ref{Fig:ms2FRegion} and \ref{Fig:gs2Pregion}
are provided in the tables ~\ref{Tab:identification} and \ref{Tab:letters} together with the resonance energies and widths of the resonances identified by Sossah~\etal~\cite{Sossah2012}. Extracting oscillator strengths requires knowledge of individual fractional abundances of Ca$^+$ ions in the ground level and the metastable levels contributing to each peak feature. Since many of the peak features are blends of several fine-structure components a detailed level-to-level assignment of peaks is not possible and hence, the oscillator strength cannot be recovered from the observed resonance strengths.

Uncertainties of the resonance strengths are directly related to the uncertainties quoted for the present cross sections, i.e., they are of the order of $\pm 30$\% and may be larger because of statistical uncertainties. Many of the Lorentzian widths provided in the tables have large relative uncertainties. This is especially true when the fitted Lorentzian widths are comparable to or smaller than the photon energy bandwidths used in the experiments.  The assignment of transitions (from a given initial level to an excited autoionizing level) to the experimental peaks is guided by the similarities seen in the experimental spectrum and the theory model spectrum displayed in figure~\ref{Fig:overviewcompnew}. The knowledge of the initial level of a given transition from theory and partly also from the existing experiments is the key to an appropriate assignment.  Similarities of the measured and calculated resonance energies and widths provide additional confidence in the level assignments inferred from the work by Sossah~\etal~\cite{Sossah2012} for the present experimental resonances.

%
%
%

\begin{longtable}[c]{cccccccccc}
\endlastfoot\hline\endfoot
 \caption{\label{Tab:identification} Identification of resonances observed in the experiments. The numbers correspond to peak labels in  figures~\ref{Fig:Eph23to28}, \ref{Fig:ms2FRegion} and \ref{Fig:gs2Pregion}.  Along with  the assignments of individual transitions, the resonance energies (in eV), the Lorentzian widths $\Gamma$  (in meV) and the resonance strengths S (in Mb\,eV) inferred from fits to the experimental resonance features are provided together with their corresponding uncertainties $\Delta$E$_{res}$, $\Delta \Gamma$, and  $\Delta$S, respectively. The experimental energies and widths are compared with those calculated by Sossah \etal~\cite{Sossah2012} (columns 2 and 3). For further details see text.  }\\
\hline\hline
no. &  E$_{th}$   & $\Gamma_{th}$ 	& transition & E$_{res}$ & $\Delta$E$_{res}$ & $\Gamma$	& $\Delta \Gamma$ & S        &   $\Delta$S  \\
	  &   eV     &	meV			&            &    eV  &	meV   		    &   meV   &      meV	  & Mb\,eV   &     Mb\,eV   \\	
\hline & & \\[-11pt]\endfirsthead
 \caption[]{(continued) Identification of resonances}\\
 \hline\hline
label &  E$_{th}$   & $\Gamma_{th}$ 	& transition & E$_{res}$ & $\Delta$E$_{res}$ & $\Gamma$	& $\Delta \Gamma$ & S        &   $\Delta$S  \\
	  &   eV     &	meV			&            &    eV  &	meV   		    &   meV   &      meV	  & Mb\,eV   &     Mb\,eV   \\	
 \hline & & \\[-11pt]\endhead
1 &	23.33 &	42.3	& $^2$D$^e_{3/2} \to (3p^53d~^3{\textrm P})4s~^2{\textrm P}^0_{1/2}$ &	23.652 &	11  &	42  &	15 &	0.100   & 0.087\\
2 &	23.44 &	40.9	& $^2$D$^e_{5/2} \to (3p^53d~^3{\textrm P})4s~^2{\textrm P}^0_{3/2}$ &	23.767 &	7   &	38  &	12 &	0.176	& 0.097\\
2 &	23.46 &	40.9	& $^2$D$^e_{3/2} \to (3p^53d~^3{\textrm P})4s~^2{\textrm P}^0_{3/2}$ &	23.767 &	7   &	38  &	12 &	0.176	& 0.097\\
3 &	24.37 &	15.4	& $^2$D$^e_{5/2} \to 3p^5(3d^2~^1{\textrm G})~^2{\textrm F}^0_{7/2}$ &	24.673 &	6   &	7	& 11   &	0.613	& 0.19\\
4 &	24.53 &	14.2	& $^2$D$^e_{5/2} \to 3p^5(3d^2~^1{\textrm G})~^2{\textrm F}^0_{5/2}$ &	24.824 &	21  &	5	& 11   &	0.501	& 0.17\\
4 &	24.55 &	14.2	& $^2$D$^e_{3/2} \to 3p^5(3d^2~^1{\textrm G})~^2{\textrm F}^0_{5/2}$ &	24.824 &	21	& 5	    & 11   &	0.501	& 0.17\\
5 &	24.76 &	17.7	& $^2$D$^e_{5/2} \to 3p^5(3d^2~^1{\textrm D})~^2{\textrm D}^0_{5/2}$ &	24.947 &	7 	& 16    &	13 &	0.256	& 0.11\\
5 &	24.77 &	17.7	& $^2$D$^e_{3/2} \to 3p^5(3d^2~^1{\textrm D})~^2{\textrm D}^0_{5/2}$ &	24.947 &	7 	& 16    &	13 &	0.256	& 0.11\\
6 &	24.84 &	119.6	& $^2$D$^e_{5/2} \to 3p^5(3d^2~^1{\textrm D})~^2{\textrm F}^0_{5/2}$ &	25.063 &	23	& 232   &	83 &    0.669	& 0.27\\
6 &	24.85 &	119.6	& $^2$D$^e_{3/2} \to 3p^5(3d^2~^1{\textrm D})~^2{\textrm F}^0_{5/2}$ &	25.063 &	23	& 232   &	83 &	0.669	& 0.27\\
6 &	24.94 &	125.1	& $^2$D$^e_{5/2} \to 3p^5(3d^2~^1{\textrm D})~^2{\textrm F}^0_{7/2}$ &	25.063 &	23	& 232   &	83 &	0.669	& 0.27\\
7 &	26.17 &	17.6	& $^2$D$^e_{5/2} \to (3p^53d~^3{\textrm D})4s~^2{\textrm F}^0_{5/2}$ &	26.474 &	7	&   8   &	11 &	3.34	& 1.1\\
8 &	26.37 &	75.3	& $^2$D$^e_{5/2} \to (3p^53d~^1{\textrm F})4s~^2{\textrm F}^0_{5/2}$ &	26.330 &	40	& 78    &	25 &	1.69	& 0.74\\
8 &	26.39 &	75.3	& $^2$D$^e_{3/2} \to (3p^53d~^1{\textrm F})4s~^2{\textrm F}^0_{5/2}$ &	26.330 &	40	& 78    &	25 &	1.69	& 0.74\\
9 &	26.95 &	2.1	    & $^2$S$^e_{1/2} \to 3p^5(3d^2~^1{\textrm D})~^2{\textrm P}^0_{1/2}$ &	26.951 &	10	& 1     &	11 & 	1.04	& 0.35\\
10&	27.00 &	1.7	    & $^2$S$^e_{1/2} \to 3p^5(3d^2~^1{\textrm D})~^2{\textrm P}^0_{3/2}$ &	27.046 &	6	& 0     &   11 &	3.09	& 0.94\\
11&	28.20 &	2.3	    & $^2$S$^e_{1/2} \to 3p^5(4s2~^1{\textrm S})~^2{\textrm P}^0_{3/2} $ &	28.203\protect\footnotemark \footnotetext{NIST energy is 28.1995~eV~\cite{NIST2016}} &	5	& 0	    & 11   &	8.93	& 2.7\\
12&	28.57 &	1.8	    & $^2$S$^e_{1/2} \to 3p^5(4s2~^1{\textrm S})~^2{\textrm P}^0_{1/2} $ &	28.545\protect\footnotemark \footnotetext{NIST energy is 28.5441~eV~\cite{NIST2016}} &	5 	& 0	    & 11   &	5.52	& 1.7\\
13&	29.24 &	251.6	& $^2$D$^e_{5/2} \to 3p^5(3d^2~^3{\textrm F})~^2{\textrm F}^0_{5/2}$ &	29.321\protect\footnotemark \footnotetext{found at 29.33~eV by Kjeldsen~\etal~\cite{Kjeldsen2002d}} &	11	& 335   &	24 &	10.9	& 3.3\\
13&	29.25 &	251.6	& $^2$D$^e_{3/2} \to 3p^5(3d^2~^3{\textrm F})~^2{\textrm F}^0_{5/2}$ &	29.321 &	11	& 335   &	24 &	10.9	& 3.3\\
13&	29.33 &	263.5	& $^2$D$^e_{5/2} \to 3p^5(3d^2~^3{\textrm F})~^2{\textrm F}^0_{7/2}$ &	29.321 &	11	& 335   &	24 &	10.9	& 3.3\\
14&	30.06 &	4.2	    & $^2$D$^e_{5/2} \to 3p^5(3d^2~^3{\textrm F})~^2{\textrm D}^0_{3/2}$ &	30.091 &	5 	& 3     &	11 &	3.17	& 0.95\\
14&	30.07 &	0.5	    & $^2$D$^e_{5/2} \to 3p^5(3d^2~^3{\textrm F})~^2{\textrm D}^0_{5/2}$ &	30.091 &	5 	& 3	    & 11   & 	3.17	& 0.95\\
14&	30.08 &	4.2	    & $^2$D$^e_{3/2} \to 3p^5(3d^2~^3{\textrm F})~^2{\textrm D}^0_{3/2}$ &	30.091 &	5   & 3	    & 11   &	3.17	& 0.95\\
14&	30.09 &	0.5	    & $^2$D$^e_{3/2} \to 3p^5(3d^2~^3{\textrm F})~^2{\textrm D}^0_{5/2}$ &	30.091 &	5 	& 3	    & 11   &	3.17	& 0.95\\
15&	30.17 &	5.6	    & $^2$S$^e_{1/2} \to 3p^5(3d^2~^1{\textrm S})~^2{\textrm P}^0_{3/2}$ &	29.993\protect\footnotemark \footnotetext{found at 29.99~eV by Kjeldsen~\etal~\cite{Kjeldsen2002d}} &	5	& 6	    & 11   &	2.24	& 0.67\\
16&	30.20 &	12.3	& $^2$D$^e_{3/2} \to 3p^5(3d^2~^3{\textrm P})~^2{\textrm P}^0_{1/2}$ &	30.203 &	6	& 15    &	11 &	1.36	& 0.41\\
17&	30.23 &	10.9	& $^2$D$^e_{5/2} \to 3p^5(3d^2~^3{\textrm P})~^2{\textrm P}^0_{3/2}$ &	30.233 &	6	& 15    &	11 &	2.18	& 0.66\\
17&	30.24 &	10.9	& $^2$D$^e_{3/2} \to 3p^5(3d^2~^3{\textrm P})~^2{\textrm P}^0_{3/2}$ &	30.233 &	6	& 15    &	11 &	2.18	& 0.66\\
18&	30.45 &	5.3	    & $^2$S$^e_{1/2} \to 3p^5(3d^2~^1{\textrm S})~^2{\textrm P}^0_{1/2}$ &	30.258\protect\footnotemark \footnotetext{found at 30.26~eV by Kjeldsen~\etal~\cite{Kjeldsen2002d}} &	7	& 5	    & 11   &    1.16   & 0.41\\
19&	30.78 &	0.2	    & $^2$D$^e_{3/2} \to (3p^53d~^3{\textrm P})4d~^2{\textrm P}^0_{1/2}$ &	30.653 &	10	&  1	& 11   &    0.195	& 0.063\\
20&	30.91 &	63.2	& $^2$D$^e_{5/2} \to (3p^53d~^3{\textrm P})4d~^2{\textrm F}^0_{5/2}$ &	30.846 &	14	& 43    & 12   &    0.733	& 0.25\\
20&	30.93 &	63.2	& $^2$D$^e_{3/2} \to (3p^53d~^3{\textrm P})4d~^2{\textrm F}^0_{5/2}$ &	30.846 &	14	& 43    &	12	&   0.733	& 0.25\\
21&	30.95 &	0.2	    & $^2$D$^e_{5/2} \to (3p^53d~^3{\textrm P})4d~^2{\textrm P}^0_{3/2}$ &	30.896 &	6	& 1	    & 11	&   0.35	& 0.11\\
21&	30.97 &	0.2	    & $^2$D$^e_{3/2} \to (3p^53d~^3{\textrm P})4d~^2{\textrm P}^0_{3/2}$ &	30.896 &	6	& 1	    & 11	&   0.35	& 0.11\\
22&	30.99 &	27.8	& $^2$D$^e_{3/2} \to (3p^53d~^3{\textrm P})4d~^2{\textrm D}^0_{5/2}$ &	30.917 &	25	& 107   &	15	& 2.69	    & 1.1\\
22&	31.04 &	96.8	& $^2$D$^e_{5/2} \to (3p^53d~^3{\textrm P})4d~^2{\textrm F}^0_{7/2}$ &	30.917 &	25	& 107   &	15	& 2.69	    & 1.1\\
22&	31.10 &	51	    & $^2$D$^e_{5/2} \to (3p^53d~^3{\textrm F})4d~^2{\textrm F}^0_{7/2}$ &	30.917 &	25	& 107   &	15	& 2.69	    & 1.1\\
23&	31.25 &	0.7	    & $^2$D$^e_{5/2} \to (3p^53d~^3{\textrm D})4d~^2{\textrm F}^0_{7/2}$ &	31.155 &	16	& 5     &	12  & 0.07	    & 0.04\\
24&	31.42 &	2.7	    & $^2$D$^e_{5/2} \to (3p^53d~^3{\textrm D})4d~^2{\textrm F}^0_{5/2}$ &	31.314 &	16	& 3	    & 12    &	0.08	& 0.09\\
24&	31.43 &	2.7	    & $^2$D$^e_{3/2} \to (3p^53d~^3{\textrm D})4d~^2{\textrm F}^0_{5/2}$ &	31.314 &	16	& 3	    & 12	& 0.08	    & 0.09\\
25&	32.02 &	12.3	& $^2$S$^e_{1/2} \to 3p^5(3d^2~^3{\textrm P})~^2{\textrm P}^0_{1/2}$ &	31.896 &	10	& 1	    & 11	& 0.285	    & 0.093\\
26&	32.07 &	10.9	& $^2$S$^e_{1/2} \to 3p^5(3d^2~^3{\textrm P})~^2{\textrm P}^0_{3/2}$ &	31.934 &	6	& 18    &	11	& 1.27	    & 0.39\\
27&	32.26 &	3.2	    & $^2$D$^e_{3/2} \to (3p^53d~^3{\textrm F})4d~^2{\textrm P}^0_{1/2}$ &	32.245 &	16	& 5	    & 17	& 0.067	    & 0.035\\
28&	32.52 &	2.3	    & $^2$D$^e_{3/2} \to (3p^53d~^1{\textrm P})4d~^2{\textrm P}^0_{1/2}$ &	32.565 &	14	& 8	    & 11	& 0.522	    & 0.21\\
28&	32.54 &	3.7	    & $^2$D$^e_{5/2} \to (3p^53d~^1{\textrm D})4d~^2{\textrm D}^0_{5/2}$ &	32.565 &	14	& 8	    & 11	& 0.522	    & 0.21\\
29&	32.63 &	42.8	& $^2$D$^e_{3/2} \to (3p^53d~^1{\textrm D})4d~^2{\textrm F}^0_{5/2}$ &	32.685 &	44	& 12    &	13	& 0.276	    & 0.12\\
29&	32.64 &	40.5	& $^2$D$^e_{5/2} \to (3p^53d~^1{\textrm D})4d~^2{\textrm F}^0_{7/2}$ &	32.685 &	44	& 12    &	13	& 0.276	    & 0.12\\
30&	32.75 &	3.5	    & $^2$D$^e_{3/2} \to (3p^53d~^1{\textrm D})4d~^2{\textrm D}^0_{3/2}$ &	32.738 &	29	& 13    &	14	& 0.198	    & 0.098\\
31&	32.86 &	3.9	    & $^2$D$^e_{3/2} \to (3p^53d~^1{\textrm D})4d~^2{\textrm P}^0_{1/2}$ &	32.840 &	61	&  1    &	17	& 0.095	    & 0.075\\
32&	33.21 &	69.7	& $^2$S$^e_{1/2} \to (3p^53d~^1{\textrm P})4s~^2{\textrm P}^0_{1/2}$ &	33.197 &    5	& 83    &	11	& 239	    & 72\\
32&	33.22 &	72.6	& $^2$S$^e_{1/2} \to (3p^53d~^1{\textrm P})4s~^2{\textrm P}^0_{3/2}$ &	33.197 &	5	& 83    &	11	& 239	    & 72\\
32 & \multicolumn{3}{c}{present 594cc DARC calculation:} & 33.224 & &  60      & &   225.8     & \\
33a&	&           & $^2$S$^e_{1/2} \to (3p^54d~^1{\textrm P})4s~^2{\textrm P}^0_{1/2}$ &	37.643 &	16	& 17    &	11	& 13.2	    & 4.1\\
33b& 	&       	& $^2$S$^e_{1/2} \to (3p^54d~^1{\textrm P})4s~^2{\textrm P}^0_{3/2}$ &	37.669 &	16	& 12    &	11	& 12.3	    & 3.8\\
33 & \multicolumn{3}{c}{present 594cc DARC calculation:} & 37.641 & & 38 & & 63.25 & \\
34a	&	&       	& $^2$S$^e_{1/2} \to (3p^55d~^1{\textrm P})4s~^2{\textrm P}^0_{1/2}$ &	39.534 &	16	& 2     &	12	& 3.31	    & 1.1\\
34b	&	&       	& $^2$S$^e_{1/2} \to (3p^55d~^1{\textrm P})4s~^2{\textrm P}^0_{3/2}$ &	39.558 &	16	& 2	    &   12	& 4.16	    &1.3\\

\hline\hline
\end{longtable}


\begin{longtable}[c]{ccccccc}
\scriptsize
\endlastfoot\hline\endfoot
 \caption{\label{Tab:letters} Parameters of unidentified resonances observed in the experiments. The letters in the first column correspond to peak labels in  figures~\ref{Fig:Eph23to28}, \ref{Fig:ms2FRegion} and \ref{Fig:gs2Pregion}.  The resonance energies (in eV), the Lorentzian widths (in meV) and the resonance strengths as inferred from fits to the experimental resonance features are provided together with the associated  uncertainties $\Delta$E$_{res}$, $\Delta \Gamma$, and  $\Delta$S, respectively.
Peaks labeled a through f are assumed to belong to a group of transitions $ ^2{\textrm D}^e_{3/2,5/2} \to (3p^5 3d~^3{\textrm F}$, $^1{\textrm D}$, $^1{\textrm F})4s~^2{\textrm F}^o_{5/2,7/2}$
predicted~\cite{Sossah2012} at energies between 25.70 and 26.02~eV. The peak labeled j is probably associated with a transition of the type $^2{\textrm D}^e_{3/2,5/2} \to (3p^5 3d~^3{\textrm P}$, $^3{\textrm F}$, $^3{\textrm D})4d~^2{\textrm F}^o$, $^2{\textrm P}^o$, $^2{\textrm D}^o$. The peaks labeled k through n are assumed to belong to a group of transitions $ ^2{\textrm D}^e_{3/2,5/2} \to (3p^5 3d~^1{\textrm P})4d~^2{\textrm D}^o$, $^2{\textrm P}^o$ predicted~\cite{Sossah2012} at energies between 32.37 and 32.44~eV.
  For further details see text.  }\\
\hline\hline
label &  E$_{res}$ & $\Delta$E$_{res}$ & $\Gamma$	& $\Delta \Gamma$ & S        &   $\Delta$S  \\
	 &    eV  &	meV   		    &   meV   &      meV	  & Mb\,eV   &     Mb\,eV   \\	
\hline & & \\[-11pt]\endfirsthead
 \caption[]{(continued) Parameters of un-identified resonances}\\
 \hline\hline
labl & E$_{res}$ & $\Delta$E$_{res}$ & $\Gamma$	& $\Delta \Gamma$ & S        &   $\Delta$S  \\
	 &    eV  &	meV   		    &   meV   &      meV	  & Mb\,eV   &     Mb\,eV   \\	
 \hline & & \\[-11pt]\endhead
a &	25.774 &	12	& 3 &	11 &	0.256 &	0.085\\
b &	25.832 &	22	& 1	&   11 &	0.173 &	0.064\\
c &	25.877 &	14	& 1	&   11 &	0.21  &	0.076\\
d &	25.921 &	14	& 45 &	13 &	1.21  &	0.42\\
e &	25.965 &	6	& 1 &	11 &	1.05  &	0.32\\
f &	26.030 &    7	& 34 &   12 &	1.42  &	0.46\\
g &	26.659 &	40	& 6	&   16 &	0.29  &	0.19\\
h &	26.869 &	33	& 4	&   16 &	0.298 &	0.18\\
i &	28.166 &	5 	& 1	&   11 &	1.97  &	0.60\\
j &	30.720 &	80	& 3	&   14 &	0.029 &	0.026\\
k &	32.322 &	7 	& 1	&   11 &	0.663 &	0.20\\
l &	32.353 &	7 	& 1	&   11 &	0.708 &	0.22\\
m &	32.408 &	33	& 1	&   11 &	0.563 &	0.21\\
n &	32.442 &	10	& 4	&   11 &	0.508 &	0.16\\
\hline\hline
\end{longtable}

\begin{figure}
\begin{center}
\includegraphics[width=\columnwidth]{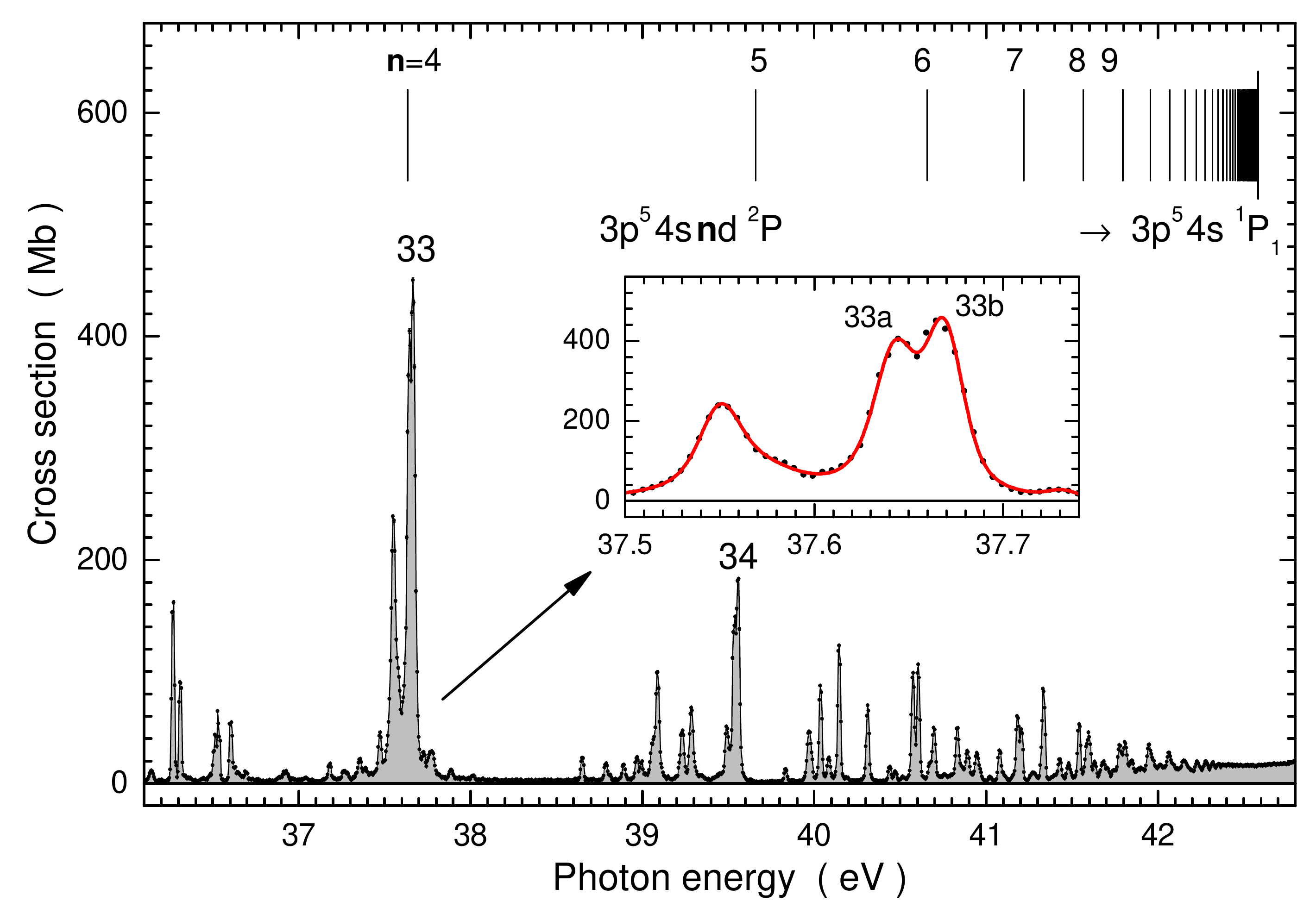}
\caption{\label{Fig:Eph36p1to42p8} Present experimental spectrum measured at 17~meV resolution in the energy region 36.1 to 42.8~eV. A Rydberg sequence of resonances is recognized which is assigned to $(3p^5 nd~^1{\textrm P})4s~^2{\textrm P}^o$ terms with $n = 4, 5, 6, ...$. The series starts with the ''giant'' resonance displayed in figure~\ref{Fig:gs2Pregion}. The energies associated with the resonances belonging to this series are indicated by vertical bars. The series limit is the $3p^5 4s~^1{\textrm P}_1$ level. The inset shows resonance 33 on an expanded energy scale. Two separate peaks (33a and 33b) are visible which may be associated with the doublet $(3p^5 nd~^1{\textrm P})4s~^2{\textrm P}^o_{1/2}$ and $(3p^5 nd~^1{\textrm P})4s~^2{\textrm P}^o_{3/2}$. The solid (red) line in the inset is a fit to the peak features using five Voigt profiles. For further details see the text.
}
\end{center}
\end{figure}

\begin{figure}
\begin{center}
\includegraphics[width=\columnwidth]{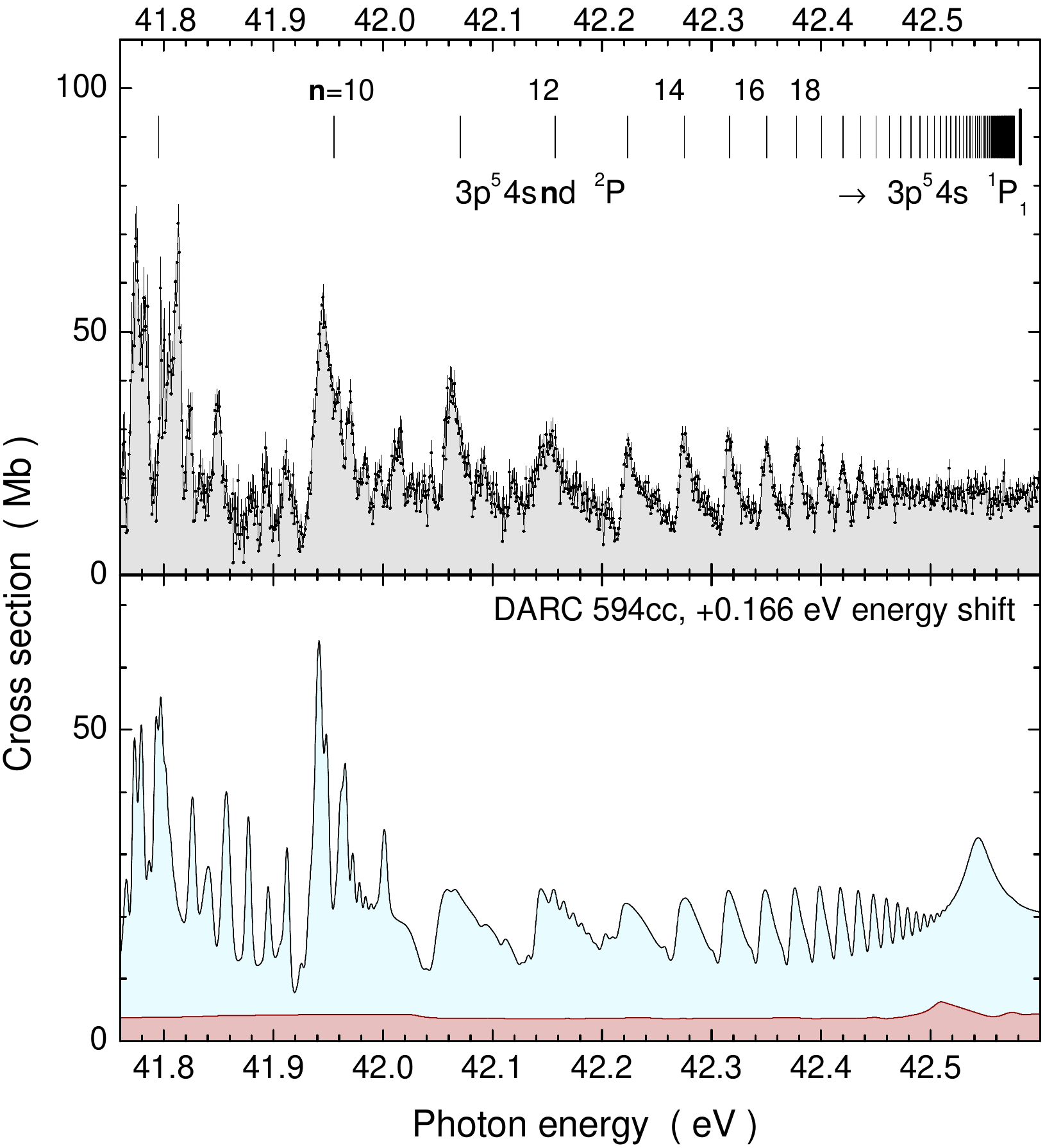}
\caption{\label{Fig:Ryd4p6}(Colour online) The upper panel shows a detail of the experimental spectrum displayed in figure~\ref{Fig:Eph36p1to42p8}. The energy resolution in this scan measurement is 4.6~meV. The Rydberg sequence of resonances assigned to $(3p^5 nd~^1{\textrm P})4s~^2{\textrm P}^o$ terms with $n = 9, 10, 11, 12 ...$ is now resolved up to the principal quantum number $n=22$. The energies associated with the resonances belonging to this series are indicated by vertical bars. The lower panel shows the associated result of the present 594-level DARC calculation. The theoretical spectrum has been convoluted with a 4.6~meV FWHM Gaussian to be comparable with the experimental spectrum. It has been shifted up in energy by +0.166~eV for easier comparison.  The Ca$^+$ ground-level contribution to the theoretical model spectrum is displayed with light (blue) shading. It has been added to the contribution arising from metastable Ca$^+$ which is displayed with darker (red)  shading (note the linear scale).
}
\end{center}
\end{figure}

Figure~\ref{Fig:Eph36p1to42p8} shows the present experimental spectrum at a resolution of 17~meV  in the energy range 36.1 to 42.8~eV. At energies near 42~eV the high-$n$ members of a Rydberg sequence are clearly evident. This observation is confirmed by a measurement with better resolution (4.6~meV) covering the energy range 41.76 to 42.60~eV (see figure~\ref{Fig:Ryd4p6}). Clearly, Rydberg resonances with principal quantum numbers $n$  up to $n = 22$ are resolved. The energies of the peaks associated with $n=11, 12, 13,...$ can be determined straightforwardly. Peaks with lower principal quantum numbers show  substantial fine structure and may be obscured by resonances from other overlapping sequences so that a determination of their energies is more difficult. The peak energies $E_n$ for $n=11,...,20$ can be fitted with a Rydberg formula considering a quantum defect $\Delta$ that is independent of the  principal quantum number $n$
\begin{equation}
\label{Eq:quantdefRyd}
E_n = E_{\infty}-  13.60569~{\rm eV}\frac{ Z^2_{\rm eff}}{(n-\Delta)^2},
\end{equation}
with $Z_{\rm eff} = 2$. Actually, it is this fit that also determines which principal quantum numbers are associated with the individual Rydberg resonances. The best fit was obtained with $\Delta = 0.6836$ resulting in
 $E_{\infty} = (42.582 \pm 0.001)$~eV. The series limit is the minimum energy required to lift one of the Ca$^+$ electrons to the continuum leaving behind a Ca$^{2+}$ ion possibly in an excited state. Since it is not known {\textit a priori} whether the series is associated with an initial ground-level or metastable Ca$^+$ ion, all the threshold energies shown as vertical bars in figure~\ref{Fig:overviewcompnew} have to be considered. These energies follow from the NIST tables~\cite{NIST2016} of levels of Ca$^+$ and Ca$^{2+}$ ions and are known with high accuracy. The threshold energy that matches the experimentally derived $E_{\infty} = 42.582$~eV is that of the $3s^2 3p^5 4s~^1{\textrm P}^o_1$ level populated by removing a $3p$ electron from ground-level Ca$^+(3s^2 3p^6 4s~^2{\textrm S}_{1/2})$. The next closest level would be $3s^2 3p^5 4s~^3{\textrm P}^o_0$ at 42.324~eV, which is 258~meV below the derived $E_{\infty}$. This is far outside the uncertainty of $\pm 5$~meV of the present energy scale and that of the fitting procedure. With the knowledge that the initial level is $^2{\textrm S}_{1/2}$ and the fact that the photon-induced dipole-allowed transitions populate solely $^2{\textrm P}^o_{1/2,3/2}$ levels it follows that a $3p$ electron is transferred to a $ns$ or a $nd$ sublevel in the first step of the photoionisation process, the more likely channel being $3p \to nd$, while $3p \to ns$ cannot be excluded. The second step is an Auger decay.

On the basis of the fit results, the sequence of Rydberg energies $E_n$ can be extrapolated towards lower $n$ values ($n < 11$). The upper panel of figure~\ref{Fig:Ryd4p6} shows that the extrapolated energies $E_{10}$ and $E_{9}$ fit rather well to peak features in the high-resolution measurement. In figure~\ref{Fig:Eph36p1to42p8} the energies $E_n$ are indicated for $n$  down to $n=4$ and, indeed, they match prominent peak features in the experimental spectrum. The relatively broad, large peak (33) with $n=4$ is undoubtedly associated with the Ca$^+(3p^5 4d~^1{\textrm P})4s~^2{\textrm P}^o$ term. The extrapolated fit of $E_n$ to $n=4$ is about 30~meV below the measured peak position. The energy deviation is understandable, considering the relatively low principal quantum number for which the quantum defect can no longer be expected to be independent of $n$. The peak with the next lower $n$ is the ''giant resonance'' with $n=3$. The associated energy from the fit is 32.439~eV with the constant quantum defect $\Delta = 0.6836$. This is about 0.76~eV below the experimental peak position. In order to match the measured resonance energy the quantum defect for $n=3$ has to be adjusted to $\Delta =  0.5919$ which is reasonable considering the low principal quantum number.

\begin{figure}
\begin{center}
\includegraphics[height=15cm]{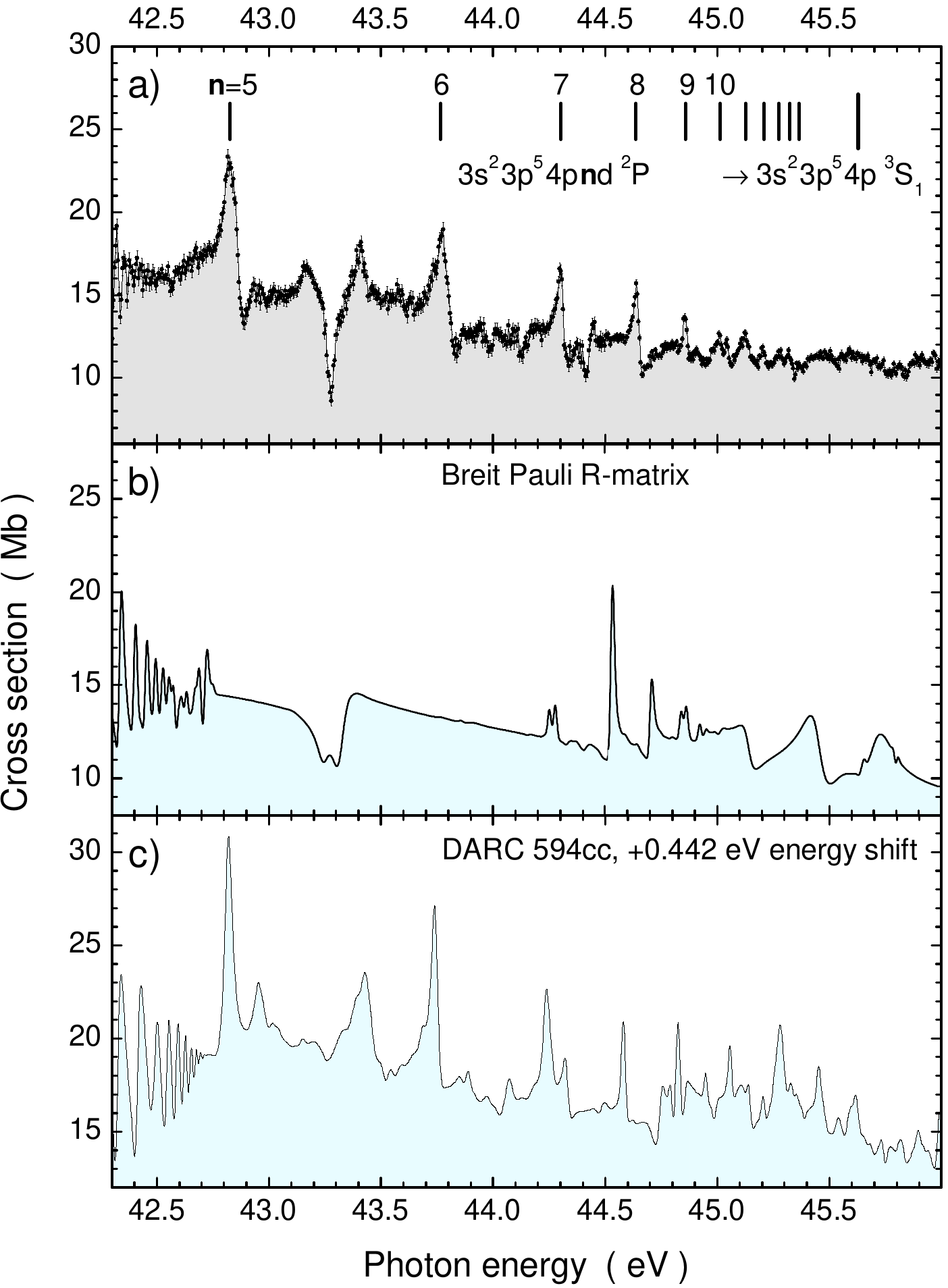}
\caption{\label{Fig:Rydzoom2}(Colour online)
 Panel a) displays the present experimental spectrum measured at 17~meV resolution in the energy region 42.3 to 46.0~eV. At least one Rydberg sequence of resonances can be clearly recognized. We assign this sequence to $(3s^2 3p^5 nd~^1{\textrm P})4p~^2{\textrm P}^o$ terms with $n = 5, 6, 7, ...$. The series limit is the $3s^2 3p^5 4p~^3{\textrm S}_1$ level. The energies of the resonances are shown as vertical bars. In panel b) the result of the Breit-Pauli {\textit R}-matrix calculation carried out by Sossah~\etal~\cite{Sossah2012} is shown in the identical energy range.  Most of the many features in the calculated spectrum cannot be related to the experimental result. However, it is pointed out that a strong adjacent pair of window resonances is predicted at about 43.3~eV where the experiment shows a clear single window resonance which appears to be part of a second Rydberg sequence with rapidly decreasing peak sizes. Such higher members of a sequence are not visible in panel b). Panel c) displays the result of the present  594-level DARC calculation. A series of peaks appears to reproduce the $(3s^2 3p^5 nd~^1{\textrm P})4p~^2{\textrm P}^o$ resonances seen in the experimental spectrum.
 The theory curves have been convoluted with 17~meV FWHM Gaussians for the comparison with the experimental data. They are dominated by the ground-level contribution to the PI model spectrum of Ca$^+$. A small cross-section contribution arising from $^2{\textrm D}$ metastable levels is outside of the present range of the cross-section axis.
}
\end{center}
\end{figure}
The lower panel of figure~\ref{Fig:Ryd4p6} displays the result of the present 594-level DARC calculation in the identical energy range. However, the theoretical spectrum was shifted by +0.166~eV towards higher photon energies in order to match the experimental peak features. The agreement of the (shifted) calculated peak positions and shapes with the experiment is remarkable. Only the resonance feature predicted near 42.55~eV is not exactly there in the experimental spectrum. This peak is rather found near 42.8~eV in the experiment (see figure~\ref{Fig:Rydzoom2}). The theoretical spectrum is composed of relatively smooth cross section contributions arising from direct PI processes and of resonance contributions both predominantly associated with  excitation of  ground-level Ca$^+$ ions.  This is in accord with the experimental assignment of the Rydberg series to $3p^6 4s~^2{\textrm S}_{1/2} \to (3p^5 nd~^1{\textrm P})4s~^2{\textrm P}^o$ transitions.

At energies beyond 42.6~eV further Rydberg series are observed in the measurement. Panel a) of figure~\ref{Fig:Rydzoom2} displays the present experimental spectrum at 17~meV energy resolution in the energy range 42.3 to 46.0~eV. The observed resonances have distinct Fano-type line profiles. Beside a pronounced window resonance near 43.28~eV, a number of resonances are present with an energy pattern of a Rydberg sequence starting with an asymmetric line at about 42.85~eV. A fit to this resonance suggests a Fano asymmetry parameter $q=-2.1 \pm 0.1$ and a natural line width of $48 \pm 4$~meV. A tentative assignment of this resonance to  a term $3s^2 3p^5 4p 5d~^2{\textrm P}$ populated by a $3p 4s \to 4p 5d$ double-excitation transition starting from ground level Ca$^+$ is in agreement with the further findings of energies along the sequence $4pnd$ with $n=6,7,8,9$. The fit of the experimental resonance energies with a Rydberg formula of the type given by equation~\ref{Eq:quantdefRyd} yields a constant quantum defect $\Delta = 0.592 \pm 0.003$ and a series limit $E_{\infty} = (45.627 \pm 0.002)$~eV. A comparison with the available photoionisation thresholds in the same manner as discussed above for the two preceding figures shows that the series limit is associated with the Ca$^{2+}(3s^2 3p^5 4p~^3{\textrm S}_1)$ level at 45.619~eV~\cite{NIST2016} which can be reached from the ground level of Ca$^+$ by removing a $3p$ electron and simultaneously exciting the $4s$ electron to the $4p$ subshell. Hence the Rydberg sequence is most likely associated with double excitations $3p 4s \to 4p nd$.

The theoretical spectra obtained from the Breit-Pauli {\textit R}-matrix approach and by the present 594-level DARC calculations are shown in panels b) and c) of figure~\ref{Fig:Rydzoom2}, respectively. The resonance features in this energy range are due to photoexcitation of ground-level Ca$^+$ ions to autoionizing levels. The calculation by Sossah~\etal shows the end of the Rydberg sequence identified from figures~\ref{Fig:Eph36p1to42p8} and \ref{Fig:Ryd4p6} near 42.5~eV. Apart from that there is little resemblance to the experimental spectrum, except for a double-window resonance feature near 43.3~eV which corresponds in energy to the distinct single window resonance seen in the experiment. The DARC calculation does not show this window resonance  but features a Rydberg sequence whose resonance energies match the experimental peak positions - provided the theoretical spectrum is shifted up in photon energy by 0.442~eV.

\section{Summary and Conclusions}
High-resolution photoionisation cross-section
measurements on the Ca$^{+}$ ion at a resolution of 17~meV were
carried out at the ALS synchrotron radiation facility. In smaller energy ranges the resolving power was substantially increased with photon energy bandpass values as small as 3.3~meV. The energy scale of the experiment was calibrated to  within $\pm 5$~meV. On the basis of the resonance identifications provided by Sossah~\etal~\cite{Sossah2012} it was possible to assign the appropriate spectroscopic notations to the most important peaks observed in the present experiment. The structures in the measured and calculated spectra are similar and resonances from theory and experiment can be associated,  although resonance energies do not match perfectly. An additional criterion for the identification of specific autoionising levels is provided by the comparison of theoretical and experimental natural linewidths of resonances.

Asymmetries in resonance cross sections have been observed in the experiments indicating interference of specific PI channels. The extraction of individual Fano asymmetry parameters, natural line widths and lifetimes from the measurements is hampered by the blending of resonance features associated with experimentally unresolvable fine-structure contributions.

When compared with the present measurements, large-scale DARC calculations  show suitable agreement over the
photon energy region investigated and reproduce the
main features in the experimental spectrum. Some minor
discrepancies are found between the theoretical results obtained from the 594-level DARC collision model
and the measurements. These are attributed to the limitations of electron-correlation
incorporated in the present 594-level collision model. Larger-scale calculations,
beyond the scope of the present investigation would likely be required to
resolve this.


\ack
We acknowledge support by Deutsche Forschungsgemeinschaft
under project number Mu 1068/10,  NATO Collaborative
Linkage grant 976362 and the US Department
of Energy (DOE) under contract DE-AC03-76SF-00098 and
grant DE-FG02-03ER15424.  B M McLaughlin acknowledges the University of Georgia at Athens for the award of a visiting research professorship and Queen's University Belfast
for the award of a visiting research fellowship (VRF).
We thank Dr A M Sossah for providing the
Breit-Pauli cross sections in numerical form and Prof S T Manson for helpful communications about their theoretical work.
M M Sant'Anna acknowledges support from CNPq and FAPERJ.
The computational work was carried out at the
National Energy Research Scientific Computing Center
in Berkeley, CA, USA and at the High Performance
Computing Center Stuttgart (HLRS) of the
University of Stuttgart, Stuttgart, Germany.
This research used resources of the Advanced Light Source, which is a DOE Office of Science User Facility under contract no. DE-AC02-05CH11231.

%
%
%
%
\section*{References}

\bibliographystyle{iopart-num}
	

\providecommand{\newblock}{}

\end{document}